\documentclass[format=acmsmall, review=false, screen=true]{acmart}

\usepackage{booktabs} 
\usepackage[ruled]{algorithm2e} 
\usepackage[T1]{fontenc}
\usepackage{lmodern}

\usepackage{listings}
\usepackage{amsmath}
\usepackage[acronym]{glossaries}

\usepackage{sidecap}
\sidecaptionvpos{figure}{t}

\usepackage[subrefformat=simple,labelformat=simple]{subcaption}

\usepackage{ctable} 
\usepackage{multirow}
\newcolumntype{L}[1]{>{\raggedright\let\newline\\\arraybackslash\hspace{0pt}}m{#1}}
\newcolumntype{C}[1]{>{\centering\let\newline\\\arraybackslash\hspace{0pt}}m{#1}}
\newcolumntype{R}[1]{>{\raggedleft\let\newline\\\arraybackslash\hspace{0pt}}m{#1}}

\usepackage{listings}
\newcommand{\emote}{{{EMOTE}}}
\newcommand{\mcec}{{{M-Enercities}}}
\newcommand{\enercities}{{{Enercities}}}

\usepackage{soul}

\usepackage[nolist,nohyperlinks]{acronym}
\begin{acronym}
\acro{HRI}{Human-Robot Interaction}
\acro{AI}{Artificial Intelligence}
\acro{ML}{Machine Learning}
\acro{WoZ}{Wizard-of-Oz}
\end{acronym}

\acmJournal{THRI}
\acmVolume{0}
\acmNumber{0}
\acmArticle{00}
\acmYear{2019}
\acmMonth{0}

\setcopyright{acmcopyright}

\begin{document}
\setcopyright{acmcopyright}
\acmJournal{THRI}
\acmYear{2019} \acmVolume{1} \acmNumber{1} \acmArticle{1} \acmMonth{1} \acmPrice{15.00}\acmDOI{10.1145/3300188}

\title{Empathic Robot for Group Learning: A Field Study}

\author{Patr\'{\i}cia Alves-Oliveira}
\orcid{1234-5678-9012-3456}
\affiliation{%
  \institution{Instituto Universit\'{a}rio de Lisboa (ISCTE-IUL), CIS-IUL, and Group on Artificial Intelligence for People and Society, GAIPS, from INESC-ID}
  \streetaddress{Av. das For\c{c}as Armadas 376}
  \city{Lisbon}
  \postcode{1600-077}
  \country{Portugal}}
\email{patricia_alves_oliveira@iscte.pt}
\author{Pedro Sequeira}
\affiliation{%
  \institution{Northeastern University, College of Computer and Information Science}
  \streetaddress{202 WVH
360 Huntington Avenue}
  \city{Boston}
  \state{MA}
  \postcode{02115}
  \country{USA}
}
\email{psequeira@ccs.neu.edu}
\author{Francisco S. Melo}
\affiliation{%
  \institution{Instituto Superior T\'{e}cnico, Universidade de Lisboa, and Group on Artificial Intelligence for People and Society, GAIPS, from INESC-ID}
  \streetaddress{IST Taguspark, Av. Prof. Dr. Cavaco Silva}
  \city{Porto Salvo} 
  \postcode{2744-016}
  \country{Portugal}
}
\email{fmelo@inesc-id.pt}
\author{Ginevra Castellano}
\affiliation{%
  \institution{Department of Information Technology, Uppsala University}
  \streetaddress{L\"{a}gerhyddsv\"{a}gen $2, 751-05$}
  \city{Uppsala} 
  \country{Sweden}}
\email{ginevra.castellano@it.uu.se}
\author{Ana Paiva}
\affiliation{%
  \institution{Instituto Superior T\'{e}cnico, Universidade de Lisboa, and Group on Artificial Intelligence for People and Society, GAIPS, from INESC-ID}
  \streetaddress{IST Taguspark, Av. Prof. Dr. Cavaco Silva}
  \city{Porto Salvo} 
  \country{Portugal}}
\email{ana.paiva@inesc-id.pt}

\renewcommand\shortauthors{Alves-Oliveira et al. (2019). Empathic Robot for Group Learning}

\begin{abstract}
This work explores a group learning scenario with an autonomous empathic robot. We address two research questions: (1) Can an autonomous robot designed with empathic competencies foster collaborative learning in a group context? (2) Can an empathic robot sustain positive educational outcomes in long-term collaborative learning interactions with groups of students? To answer these questions, we developed an autonomous robot with empathic competencies that is able to interact with a group of students in a learning activity about sustainable development. Two studies were conducted. The first study compares learning outcomes in children across $3$ conditions: learning with an empathic robot; learning with a robot without empathic capabilities; and learning without a robot. The results show that the autonomous robot with empathy fosters meaningful discussions about sustainability, which is a learning outcome in sustainability education. The second study features groups of students who interact with the robot in a school classroom for two months. The long-term educational interaction did not seem to provide significant learning gains, although there was a change in
game-actions to achieve more sustainability during game-play. This result reflects the need to perform more long-term research in the field of educational robots for group learning. 
\end{abstract}


%
%

\maketitle
\section{Introduction}%
\label{Sec:Intro}

Learning is an intrinsic human ability. We learn both from our own experience and from our peers. However, learning is not just about improving performance or assimilating new knowledge. It is also about analyzing new situations, understanding different perspectives, using knowledge to find commonalities between distinct situations, discussing, and even become competent at solving conflicts. Social learning, whereby we learn with and from our peers, teachers, parents and others, plays a fundamental role in most of these broad forms of learning \cite{piaget2013play}. It is not surprising then that the educational environment stimulates social collaboration between peers during learning. Collaborative learning has been associated with improving attitudes towards school, fostering achievement, developing thinking skills, promoting interpersonal and intergroup relations \cite{blumenfeld1996learning}. 

However, as technology evolves, so do learning paradigms. Currently, technology offers novel learning tools that complement the classical learning paradigms. Massive open online courses (MOOCs) \cite{li2014watching} and intelligent tutoring systems (ITS) \cite{nwana1990overview} are just two illustrative examples of new learning tools that are available through technology. Another change in the teaching and learning paradigm comes from serious games \cite{ritterfeld2009book}. Serious games are designed for a purpose other than pure entertainment. They have been around since at least the 1950s, and their applications in education are well-documented \cite{degloria2014serious}\footnote{Interestingly, research suggests that while serious games may be more effective in terms of learning, they are not always more motivating than conventional instruction methods \cite{wouters2013meta}.}. Collaborative serious games, in particular, combine the advantages of serious games with social learning, and some studies have suggested that they support learners in articulating the knowledge that would otherwise have remained intuitive \cite{van2011learning}. Although some researchers questioned the effectiveness of collaborative serious games, research in this area is still scarce and often ambiguous \cite{wouters2013meta}. 

In this paper, we describe the use of a {\em robotic tutor in the context of a collaborative serious game}. In particular, the research described herein seeks to combine the positive aspects of intelligent and peer tutoring, such as personalization and adaptation to the learner, with collaborative serious games, such as social learning and gamification. This offers a new role for robots as technological tools in education.

Previous studies have explored the use of artificial robotic tutors and peers in both classroom environments \cite{kanda2004interactive,tanaka2007socialization, miyake2012robot} and as entertainment partners \cite{michalowski2007dancing}. However, such use is often limited to one-robot-one-user interactions, which significantly limits the social component of learning. By adopting a collaborative serious game as the interaction domain, we move from such typical one-robot-one-user interaction to a much richer scenario that involves one robot and two users, thus, a group interaction. In addition, some studies have investigated the effect of long-term interactions between children and robots regarding the quality of the relationship \cite{westlund2018measuring}, and other variables \cite{leite2013social}; however, few studies have looked at the impact of long-term interactions and their effects on learning gains, which is also a contribution of this paper.

Our ambitious application scenario poses several technological challenges, particularly in endowing the robot with a behavior that is both socially plausible and able to successfully accomplish the pedagogical goals of the activity. Our robotic tutor must be both socially competent and successful in teaching. It should also be able to perceive the difficulties faced by the students, not only from their explicit behavior but more importantly from their implicit behavior. It should be able to understand their individual affective state and the ``emotional climate'' between them to intervene in a way that is adequate. In other words, a robot should be able to {\em empathize} with the human users, individually and as a group.

A recent survey discussed the importance of empathy for an artificial agent (robot or virtual) that interacts with humans \cite{paiva2017empathy}. In the context of education, the empathy of the teacher towards the students was also shown to impact the learning process and outcomes \cite{feshbach2009empathy}. In a meta-analysis conducted by \citeN{cornelius2007learner}, empathy was one of the variables that if present in teachers is associated with positive students outcomes. In a subsequent meta-analysis conducted by \citeN{roorda2011influence}, empathy was included as one of the variables in the teacher-student's relationship that is associated with their engagement and achievements in school. This demonstrates the importance of empathy in tutor-student relationships and how it influences students' educational outcomes. However, few studies have measured this impact on learning outcomes. This paper describes an autonomous and empathic robot tutor that interacts with multiple learners via collaborative serious game and investigates the impact, both immediate and long-term, of the robot on the student's learning performance. Two field studies were performed, leading to a deeper understanding of empathy in learning and the long-term effects of having an autonomous robot in school for educational purposes. 

\subsection{Contributions}
The work presented herein was developed in the context of the \emote{} project\footnote{{EMOTE project: \url{http://www.emote-project.eu}}, which stands for EMbOdied Perceptive Tutors for Empathy-based learning project.}. We investigated the use of an {\em autonomous and empathic robot tutor for} \emph{collaborative group learning}. We explored the impact of a robot interacting and teaching groups of students in their classroom in relation to learning gains.
The contributions of this paper include the following:

\begin{enumerate}
\item We conducted a short-term evaluation study to investigate the immediate impact of the empathic robot in fostering collaborative learning.
\item We conducted a long-term term study to investigate the long-term impact of the empathic robot in terms of learning.
\item We deployed an autonomous robot in the context of real-world classrooms for group learning.
\end{enumerate}

The road-map of this work starts with section \ref{ch:state_art} presenting the state of the art regarding robots in education, groups of humans and robots, and empathy in social robots; section \ref{ch:collaborative_learning_activity} details on the collaborative learning activity, including the learning goals and the game-play dynamics; section \ref{ch:robot_tutor_behavior} refers to the development of the educational and social behaviors for the robotic tutor; section \ref{ch:method} explains the study method, hypothesis, measures, and materials used; section \ref{ch:short_term_study} describes the short-term study and Section \ref{ch:long_term_study} the long-term study; section \ref{ch:conclusion} presents the general discussion and conclusion, including limitations and future work.

\subsection{Our previous work}
This paper focuses on the evaluation of the robotics system with respect to learning gains in students. The details of the design and implementation of the robot are not a main contribution of this paper and can be found in previous publications. In particular, the robot behaviors and the development of the empathy module of the autonomous robot are detailed in \citeN{alves2015ec} and are summarized in section \ref{Subsec:EC} of this paper. The development of the collaborative \ac{AI} that sustains the robot's game-playing and pedagogical decision-making abilities is detailed in \citeN{sequeira2015ai} and is overviewed here in section~\ref{Subsec:GameAI}. The educational dialogue dimensions for collaborative learning designed for the robot are detailed in \citeN{alves2014towards} and summarized in section \ref{Subsec:Hybrid}. Finally, the design, development, and validation of the ``Restricted-perception Wizard-of-Oz'' methodology, which allowed the successful development of the social and educational behaviors of the autonomous robot, are detailed in \citeN{sequeira2016method}, and we summarized the main steps in section~\ref{Subsec:Restricted-woz}\footnote{The source code of all the robot's components is publicly-available and can be retrieved at: \url{https://github.com/emote-project/Scenario2}. Further details on each component can be found in the ``Downloads / Components'' and ``Publications / Deliverables'' sections of the project's website at: \url{http://www.emote-project.eu/}.}.

\section{State of the art}
\label{ch:state_art}
The experiences we have during our childhood shape to some extent the way we think, grow, feel, and behave. It is thus important to surround children with nurturing and safe learning environments. The way children learn, however, is being transformed with new technologies for education, such as computers and tablets, that are enhanced with serious games \cite{savage1990conceptual} or computer-supported learning activities, to foster learning acquisitions. Robots in particular hold promise to facilitate learning outcomes and promote enjoyment during learning \cite{kennedy2015comparing}. Recent events, such as the R4L (or Robots 4 Learning) series of workshops\footnote{The workshops and special issue can be found in \url{https://r4l.epfl.ch/}.} are an example of the interest that this area is capturing in research and the potential that robots have in education. In this paper, we present research that advances the state of the art in the area of social and empathic robots for education, in particular for collaborative group learning scenarios. Thus, we now provide an overview of these areas to contextualize the contributions of the paper.

\subsection{Robots in education}
Educational robots are a subset of educational technologies in which robots are used as platforms or tools for students' learning, usually in subjects such as maths, problem solving, chemistry, etc. The use of robots as a medium to learn and understand curricular subjects started in the 60s with Papert's work in which he introduced the concept of ``robots as manipulatives'' (or physical objects) that are specifically designed to foster learning \cite{papert1980mindstorms}. An example is the LEGO Mindstorms\textregistered, which are used to teach STEAM-related curricular topics, providing a new tool for education \cite{hendler2000robots, khine2017robotics}. Furthermore, as robot technology matures, robots can be used as social actors in a classroom as a way to deliver educational content, instruct, foster discussion, challenge, scaffold, and support the learning of children in a social-intelligent manner. In fact, reviews on the applicability and potential of robots in education shows that robots are being developed for children across different learning domains \cite{mubin2013review, belpaeme2018social}. Despite the potential of robots in learning, a systematic review has shown that to have an impact on learning gains, they need to be skilfully used by teachers to attend to the students' needs. Otherwise, learning gains are not visible \cite{spolaor2017robotics}.

There is investment in the field of \ac{HRI} to develop robots that are used for learning. Pioneering research  by \citeN{kanda2004interactive} features ROBOVIE, a robot for  teaching English to Japanese children in an elementary school context. This was one of the first field studies in educational \ac{HRI}, conducted over a period of 18 consecutive days in a school. The results showed that children who exhibited a lot of interest during the starting phase had a significantly elevated English score, and the robot indeed acted as a motivational factor for learning in these cases. In South Korea, the IROBI robot was also endowed with didactic content to support young children with learning English as a second language \cite{han2008comparative}. The robot was placed in a class with very young children over a long period of time. In the same application domain, the EU H2020 L2TOR project\footnote{L2TOR project: {\protect\url{http://www.l2tor.eu/}}.} aims to study if robots can be used as tutors to support teaching preschool children a second language \cite{kennedy2016social}. A review of the applicability of robots for second language acquisition was performed by \citeN{chang2010exploring} in which they have characterized the existing robots used and the instructional media explored for this type of task. Other projects with robots for children have dedicated attention to the investigation of robots as tools for learning and supporting positive interactions, such as the Socially Assistive Robotics, an NSF Expedition in Computing project\footnote{Socially Assistive Robotics, an NSF Expedition in Computing project: {\protect\url{https://robotshelpingkids.yale.edu/}}.}.

Robots have also been used to support handwriting abilities using the paradigm of ``learning-by-teaching'' \cite{frager1970learning} in which children act as tutors of the robot, providing it with feedback for a better writing performance. This paradigm is known to benefit children's self-esteem, provide practice with hand-writing without noticing, and provide engagement in a so-called ``prot\'{e}g\'{e} effect'', which is a sense of responsibility over the robot's performance (since they are instructed to be the robot's teachers) \cite{chase2009teachable}. For example, \citeN{tanaka2012children} conducted a 6-day field trial with young students and a robot in school, and the results showed that a robot can help children efficiently learn new English verbs, when children give instructions to the robot. Additionally, projects such as the Co-writer project\footnote{Co-writer project: \protect\url{http://chili.epfl.ch/cowriter}.}, have studied the role of students as teachers of the robot. A study in the scope of the Co-writer project was conducted in school in which a group of children gathered around the NAO robot to provide feedback on its mistakes as the robot improved according to children's feedback \cite{lemaignan2016learning}. While providing feedback to the robot, children are put into small groups who are responsible for teaching the robot to write better. In a study in which the NAO robot acted as a student that needed to learn how to write and the children served as teachers who helped it write better, results showed high levels of commitment and engagement from children embracing this task \cite{hood2015children}. This demonstrated the promising results of using this educational system.
Another study investigated interpersonal distancing of children both towards a human adult or a robot facilitator within a collaborative activity \cite{chandra2015can}. The scenario involved two children performing a collaborative learning activity following a learning-by-teaching approach, which included writing a word/letter on a tactile tablet. The study showed that children felt more responsible and provided more corrective feedback when the robot was present than when a human mediator was present (replacing the robot). This suggests that the role a robot can play can have an impact in the type of interactions that emerge, particularly corrective feedback. 

Beyond the typical curricular activities, social robots are also being used for social and emotional learning. The research by \citeN{jimenez2014effect} described a study in which the behavior of a robot prompts constructive interaction with a learner when compared with two students learning the same task. The types of prompts and interactions built into the robot as it learns together with the children lead to better performance of the robotic condition. Another study focused on fostering children to develop a growing mindset as part of social and emotional training \cite{park2017growing}. Therefore, in a scenario featuring a child playing puzzle games with a robot, a fully autonomous robot was built with the capability to exhibit ``behaviors suggestive of it having either a growth mindset or a neutral mindset'' \cite{park2017growing}. The results of a study that compared two types of robots in the same scenario have shown that children who played with a growth mindset robot self-reported having a stronger growth mindset and tried harder during the task.

In a long-term study, \citeN{serholt2018breakdowns} investigated interaction breakdowns between children and a robot in a learning task. In this individual learning scenario, breakdowns in the interaction were associated with ``the robot's inability to evoke initial engagement and identify misunderstandings, confusing scaffolding, lack of consistency and fairness, and controller problems.'' In another long-term study in which a robot acted as an agent for learning, \citeN{jones2018adaptive} concluded that if a robot provides personalized and adapted scaffolding to students based on their learning, they can better regulate their own learning  \cite{jones2017know}. Furthermore, in a study of two consecutive weeks in a school, a peer-robot that exhibited behavioural personalization was found to have a positive influence on learning when interacting with children, compared to a robot that exhibited non-personalization. Personalization of the robot was defined in terms of non-verbal behaviour, linguistic content, and performance alignment. Specifically, the results from this study showed that children exhibited significantly increased learning only in the novel learning task in the personalized condition \cite{baxter2017robot}. Although these three scenarios present themselves as extremely rich and challenging for social learning, especially because a robot was deployed in school for long-term educational gains, they were built for one-robot-one-student interactions. This paper goes beyond the current state of the art by exploring collaboration in a group context using a robot designed to support learning. In addition, we evaluated the learning gains, an aspect rarely measured in studies that use robots for education. In fact, usually studies evaluate other variables, such as likability and engagement between children and robots.

\subsection{Groups of humans and robots}
As mentioned, the majority of the application scenarios developed thus far to study humans and robots are designed for one-on-one interactions in which one robot interacts with one person. Even in scenarios in which the robot is placed in a classroom with many children, the type of interactions often designed consider one-on-one interactions, e.g., \citeN{belpaeme2018guidelines}. For this study, we were interested in scenarios using groups of two or three students who can learn together with the support of a social robot. According to \citeN{du2016impact}, dyads and triads are considered groups, with dyads being considered the smallest type of a group if they share common and dependent elements. In the case of our research, dyads of students share the same learning context as part of the group.

Groups in \ac{HRI} can be studied according to different perspectives, such as (1) groups of humans interacting with one robot, (2) groups of robots interacting with one human, or (3) groups of humans interacting with groups of robots. In addition to workshops organized regarding this topic \cite{jung2017robots}, relevant studies have been conducted. For example, a study by \citeN{fraune2015rabble} showed that different number of robots (namely, a single robot or a group of robots) and the type of robots (anthropomorphic, zoomorphic, or mechanomorphic), determine the attitudes, emotions, and stereotypes that people hold when interacting with them, with anthropomorphic robots in groups being one of the preferred choices. In a field study at a university, \citeN{fraune2015three} studied how participants respond when robots (individually and in groups) enter their common space. The results showed that although participants reported enjoying interacting with both individual and robots in groups, they interacted more with groups of robots. In addition to this, the characteristics of social robots can affect the interactions. For example, entitative groups of robots (robots designed to look and act similar to each other, such as those that have a similar appearance and a shared goal) compared to single robots, were evaluated more negatively \cite{fraune2017threatening}. In a another study, the frequency for which robots acted as moderators affected the social and task-related features, namely group cohesion and task performance, in multi-party interactions \cite{short2017robot}. These studies show that a robot's behavior in a group should be carefully designed as their behavior and appearance influence interactions.

Nonetheless, the perception of robots in groups not only depends on the robots' behaviors but is also influenced by the characteristics of the people. A study conducted by \citeN{correia2018choose} showed that when people only observe robots (before any direct interaction with them), they tend to choose robots that exhibit relationship behaviors (e.g., robots that foster a group climate) over competitive robots (e.g.,  robots that are more focused in succeeding and wining). However, after a direct interaction, the results seem to change, with participants who were competitive preferring a more competitive robot, and participants who were not as competitive preferring robots with relationship-driven characteristics. This study reflects that membership preferences between groups of humans and groups of robots goes beyond robot's characteristics, extending to the characteristics of each person. In another study regarding group membership, \citeN{sembroski2017he} found that in-group and out-group perception between humans and robots can lead to more conformist behaviors from people, depending on the request type and level of authority. A different study investigated a robot's potential to shape trust in groups of humans, concluding that the robot that exhibited vulnerable behavior (in comparison with a neutral robot) promoted more group engagement and social signs of trust, with people providing more support in times of failure \cite{strohkorb2018ripple}. Furthermore, a study dedicated to the investigation of non-verbal behavior between a robot and multiple people concluded that the gaze of the robot influences people's perception of the motion of the robot and that, in turn, affects the perception of the robot's gaze \cite{vazquez2017towards}. Additionally, when incorporating robots in unstructured social environments, such as in shopping malls or crowded city streets, it is important for the robot to exhibit adequate motion for collision avoidance to navigate fluently between groups of people \cite{mavrogiannis2018social}. Therefore, considering mutual dependency between the behavior of robots and the behavior of the humans is important when designing and evaluating robotic systems aimed to be social in group contexts.

In relation to educational contexts and groups, \citeN{strohkorb2015classification} used interactive role-playing to help children improve their emotional intelligence skills. By focusing on finding ways to analyze non-verbal and verbal behavior to detect if a child was high or low in social dominance, the scenario featured groups of children, and the robot helped them in their social relations. Despite being a group scenario, the collaborative nature of the task was not the focus of the work. In fact, according to \citeN{dillenbourg1999you}, collaborative learning pertains to a situation in which particular forms of interaction among people are expected to occur that trigger learning. In robots applied to education, social robots can play roles, such as of a peer (learning together with a group of students); a tutor (teaching a group of students); a facilitator/mediator (mediating the learning interactions and interventions and helping the group to follow productive learning); a  supervisor (supervising the work being done by students and providing feedback); or even a friend (supporting the students emotionally) \cite{zaga2015effect, alves2016role, broadbent2018could}. This demonstrates the richness of interactions that can be explored in the educational domain when using robots to foster learning. Although there is a wide variety of work in \ac{HRI} exploring robots as tutors in a classroom context, thus far, the type of activity and interactions established with robots have been fundamentally individual interactions.

\subsection{Empathy in social robots}
Empathy is an essential ingredient for any successful relationship. When attempting to define empathy, we come across various definitions. Empathy has been related to affective responses towards others (affective empathy) and a cognitive understanding of the emotional state of others (cognitive empathy). \citeN{hoffman2001empathy} brought attention to the processes that underlie empathic responses, defining empathy as the psychological process that leads to feelings that are congruent to the other person's situation more than to one's own. From a more behavioral perspective, \citeN{davis2018empathy} associated empathy with the responses of someone to the experiences of another person. Holistically, \citeN{preston2002empathy} considered empathy as a concept that relates sympathy and emotional contagion. The expression of empathy is foundational with respect to interpersonal relationships and with our ability to communicate. Indeed, it is what connects us emotionally and intellectually.

In an educational context, there is a general understanding from teachers that empathy promotes positive interactions, a more supportive classroom climate, and enhances student-centered practices \cite{mcallister2002role}. In a meta-analysis conducted by \citeN{cornelius2007learner}, empathy was one of the variables associated with positive students outcomes when present in teachers. When robots are placed as collaborative partners within a group, features such as their ability to communicate and relate become relevant. Interactions with robots that are more open-ended and involve a high degree of  collaboration that must be natural and engaging; thus, empathy needs to be addressed and modeled in those robots \cite{paiva2017empathy}. Recent research has been devoted to the implementation of empathy in robots in diverse application areas, such as in health care with socially assistive robotics \cite{tapus2007emulating}. In these cases, robots are perceived as having more empathy if they provided appropriate answers during a dialogue \cite{riek2010my}. In this line of research, robots that display accurate empathic behaviors enable the perception of a closer relationship with a human \cite{cramer2010effects}. Additionally, robots with empathy were perceived as being friendlier during an entertaining scenario \cite{leite2013influence}.

Empathy also plays a role in child-robot interactions. In a study conducted by \citeN{leite2014empathic}, they explored whether a robot with empathic capabilities can influence the perception that children have towards the robot, specifically social presence, engagement and support. Indeed, in a chess game scenario, the robot that displayed empathic behaviors was perceived positively by the children, and the perceived levels of social presence, engagement and support were stable and high during a long-term interaction \cite{leite2014empathic}. This research was important for the study of robots with empathy and children in real-world settings. In a broader view of empathy in robots, \citeN{paiva2018towards} dedicated a book chapter on the creation of more humane machines in which successful interactions between humans and robots is associated with robots that are endowed with emotional processing and responses. Our research goes beyond the current state of the art as it considers how a robot can model empathy in a group setting in a collaborative learning activity.

\section{Collaborative Learning about Sustainable Development}
\label{ch:collaborative_learning_activity}

A well designed scenario for collaborative learning increases the probability that some  positive interactions occur,  thus leading to learning experiences \cite{dillenbourg1999you}. The design of the scenario for collaborative learning is of utmost importance since the interactions between the members are crucial. As stated before, thus far social robots have been used extensively in domains in which the robot is able to establish one-on-one interactions with the learner, thus supporting problem solving, personalization, feedback and scaffolding. However, in  collaborative learning, other elements come into play, such as perspective taking and understanding the consequences of actions. For the scenario in our studies, we used a collaborative multi-player game targeting issues of sustainable energy consumption. This game, which we shall refer to as \mcec{} (for Multi-player Enercities), corresponds to a multi-player, collaborative version of the original game \enercities{} \cite{knol2011enercities} (see an image of the game's interface in Fig.~\ref{Fig:enercities}). 
\subsection{Learning goals}
The game was adjusted to the current schools' curriculum and was easily integrated into the school activities, allowing for 3 children to play the game, two children and a teacher, or two children and a robot (see Fig~\ref{Fig:scenario} for an overview of the group learning setting). The group of three work together to build a sustainable city, thus learning about sustainable development.  This game is in line with Constructivism's basic idea that knowledge and meaning are built during social interaction and cooperation \cite{steffe1995constructivism}. As a collaborative group learning activity, \mcec{} motivates players to discuss and decide together how to build a sustainable city, providing exactly the type of social interaction dynamic to foster communication and cooperation as means to learn about sustainability. In fact, according to scholars, sustainable education is all about having meaningful discussions in which multiple perspectives and trade-offs are exchanged by expressing personal values \cite{fior2010children, antle2014}. In light of how sustainable education should be taught, the \mcec{} game supports learning goals related with factual knowledge about sustainable development, raised awareness for trade-offs and multiple perspectives concerning sustainability, and the role of personal values. These learning goals are elaborated in Table~\ref{tab:measure1}.

\begin{table}
\caption{Learning goals supported by the collaborative group learning game, \mcec{}.}
\label{tab:measure1}
  \renewcommand{\arraystretch}{2}
  \begin{tabular}{ L{70pt} L{300pt} }
    \hline
     \textbf{Learning goal 1} & \emph{Factual knowledge} about different energy sources\\
     \textbf{Learning goal 2} & Raise awareness about \emph{trade-offs} and the existence of multiple and possibly conflicting \emph{perspectives} in relation to sustainable development\\ 
     \textbf{Learning goal 3} & Raise awareness of \emph{personal values} in relation to sustainable development\\
    \hline
  \end{tabular}
\end{table}

\subsection{Game-play}
The game is structured in levels that players need to master to continue their city. For example, to proceed to the second level in the game, players must make the population of the city grow to a certain amount by building residential areas. However, at the same time, if the city runs out of non-renewable resources, players may get into trouble because the city is not sustainable. Additionally, the game-play is based on a turn-taking dynamic where, at each each turn, one player decides what to do and the rest of the group assists in this decision process, discussing how they would like to build their city. In order for them to decide, they have to take into consideration the city indicators and how their actions influence the sustainability of the city. Each decision can have positive and negative effects on the ``environment'', ``economy'' and citizens' ``well-being'', which is indicated by a score for each of these factors. Players can choose among a set of actions that advances the game:

\begin{figure*}[!tb]
   \centering
   \begin{subfigure}{0.45\columnwidth}
   		\centering
       	\includegraphics[width=\columnwidth]{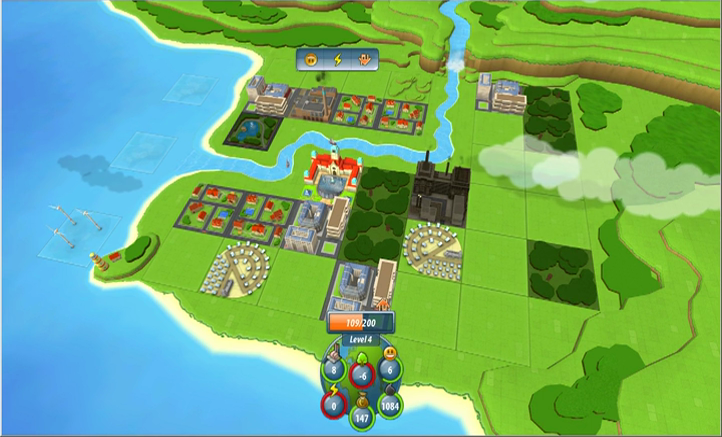}
		\caption{}\label{Fig:enercities}
   \end{subfigure}\hspace{20pt}%
   \begin{subfigure}{0.48\columnwidth}
   		\centering
       	\includegraphics[width=\columnwidth]{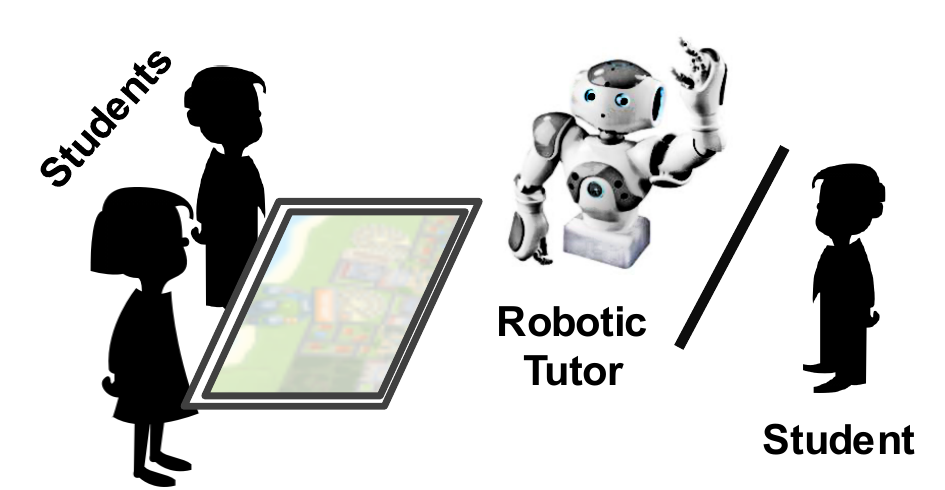}
		\caption{}\label{Fig:scenario}
   \end{subfigure}
   \caption{Learning activity used in this study: \subref{Fig:enercities} a screenshot of the \mcec{} game used in the study, which is a multi-player, turn-based, collaborative version of the \enercities{} game about sustainable development; \subref{Fig:scenario} the learning space shows the interaction of either two students with a robotic tutor or three students playing the \mcec{} game over an interactive touch-table.}
   \label{Fig:activity}
\end{figure*}

\begin{description}
	\item[Making a construction --] Players can build a construction, such as a park, a market, or an industry. The different constructions influence the city indicators differently. Thus, a park positively affects the environmental indicator of the city, an industry positively influences the economy, and a market positively influences the well-being of the citizens. However, at the same time, these constructions can have negative effects on other indicators. Players can also invest in the city's energy by building energy supplying constructions (e.g., windmills) or increase the city's size and thereby the population by constructing more residential areas.
    \item[Performing an upgrade --] Players can invest in making an upgrade on previously created constructions. They can, e.g., add a recycling facility in an already built industry to upgrade it and improve the environment without wasting additional resources. Depending on the upgrade performed, it can influence different indicators.
    \item[Applying a policy --] Players can implement new policies in their city, such as implementing an energy education program. Policies are applied in the city-major building as a way to represent how the real world functions.
    \item[Skipping a turn --] Players can choose to skip a turn for the constructions and policies to come into full effect. When skipping a turn, constructions, upgrades, and policies present in the city augment their effect, e.g., if the players have chosen to build an industry to increase the economy of the city, the city will be richer when more turns have passed. Consider turns passing as time passing (or as a time indicator). Therefore, the more the players skip turns, more benefits over previous constructions, upgrades, or policies will occur.
\end{description} 

To support turn-taking and to make every member of the group participate, there are virtual buttons located at different sides of the table, allowing players to perform different actions in turns. Since each decision can have both positive and negative effects, players must be aware of the trade-offs and the impossibility of creating a perfect city without any sacrifices. After reaching a group decision, the player closer to each button presses it, making a game action on their city.

\section{Robotic tutor}
\label{ch:robot_tutor_behavior}
To fulfill its role as an autonomous empathic tutor in the collaborative \mcec{} game, the behavior of our robot can be seen at two distinct levels: an \emph{activity-related} level and an \emph{social interactive} level. Activity-related behaviors concern the decisions of the robot in the pedagogical activity itself. In our case, it involves all game actions for \mcec{} and the necessary game-play adaptation mechanisms. The social interactive behaviors involve all verbal and non-verbal expressive behaviors of the robot during interactions with the group of students, such as the dialogue management and physical animations. Both types of behaviors are selected based on information about the state of the game and contextual information about the physical environment provided by different audio-visual capture devices. Given the intended empathic nature of the robot, its behavior also depends on the individual and collective affective state of the students. This section overviews the design and implementation of the behavioral components of the robot, including task-specific behaviors, empathic behaviors, adaptation to the students, and behaviors driven by the learning goals.

\subsection{Architecture overview}
\label{Subsec:Architecture}
\begin{figure}[!tb]
   \centering
   \includegraphics[width=0.95\textwidth]{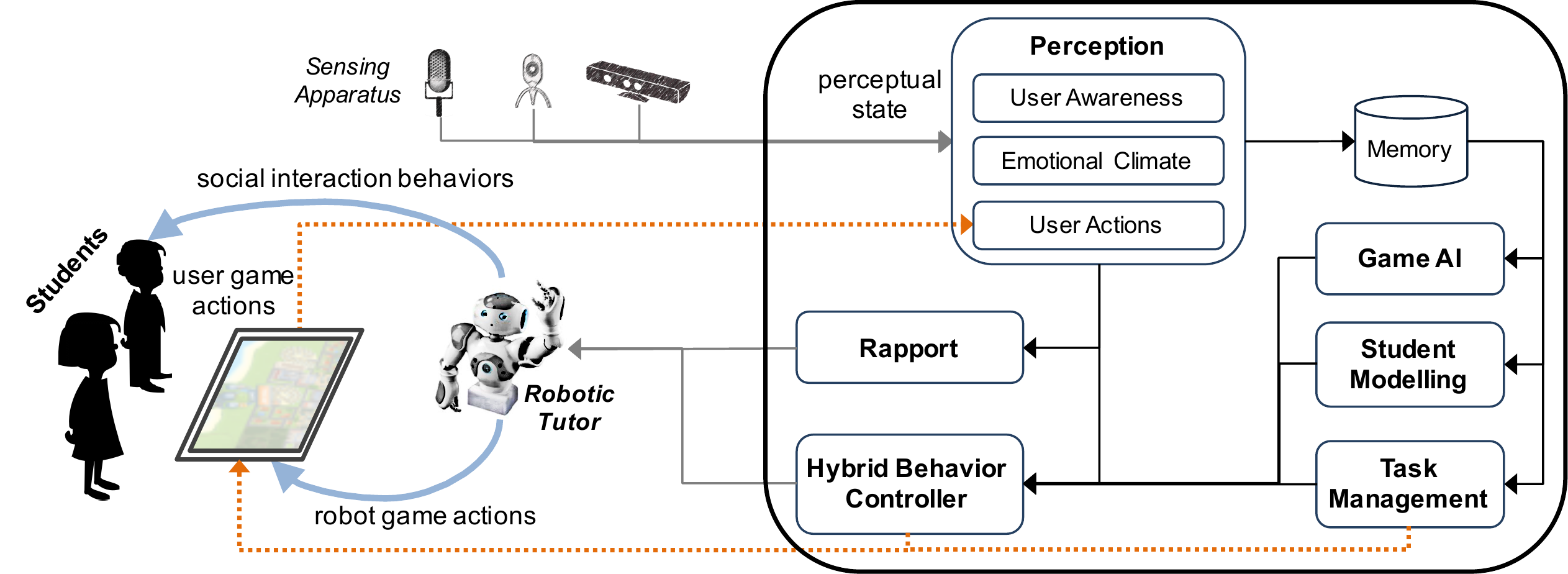}
   \caption{The overall robot system architecture implemented to autonomously control the robotic tutor during the interaction with the students in our studies. Light-blue arrows denote the robot's actions, dotted orange lines represent communication between the system and the \mcec{} game application.}
   \label{fig:architecture}
\end{figure}

The overall architecture supporting the robot used in our studies is depicted in Fig.~\ref{fig:architecture}. The system comprises the following main modules\footnote{Technical details about each component can be found at \href{http://www.emote-project.eu/components}{www.emote-project.eu/components}.}:
\begin{description}
	\item[Perception --] Module responsible for processing all information from the robot's environment, including information on the student's emotional state, social behavior and actions performed during the learning activity. It comprises three main components: the \emph{User Awareness} component processes the students' state and actions; the \emph{User Actions} component manages the students' \mcec{} game input, providing the robot with the necessary task-related context for the interaction; finally, an \emph{Emotional Climate} component is responsible for detecting group-level emotional state.
	\item[Memory --] This module keeps track of past events, and is subdivided in \emph{recent} and \emph{past event} memories. \emph{Recent Event Memory} stores the recent actions of the students in the game, collected directly from the User Actions perception component. Long term interaction, however, requires the robot to track information from previous sessions. For this reason, the information stored in the recent memory is moved to the \emph{Past Event Memory} at the end of each session.
	\item[Task Management --] This module manages the execution of the learning activity itself, e.g., starting and ending the activity and specifying which students are interacting.
	\item[Student Modeling --] This module is responsible for collecting information regarding the student's performance during the game session, using it to track possible changes in the learning and emotional state of students. Such information is used by the robot in later sessions to address specific learning challenges. 
	\item[Rapport --] This module regulates the robot's rapport during the interaction with the students. E.g., it adjusts the robot's speech volume to the student's to ensure a smooth communication; it shifts the robot's gaze towards the active speaker to provide a more natural interaction; it interrupts the robot's speech when a student is speaking; and performs back-channeling behaviors after the users' responses.
	\item[Game \ac{AI} --] This module manages all the robot's actions in the \mcec{} game. It uses the student's past actions (retrieved from Memory) to learn the strategies being used by them during the activity, and generates possible actions according to the current state of the game.
	\item[Hybrid Behavior Controller --] This is the core module controlling the robot's social interaction behavior. It decides which actions the robot should play in the \mcec{} game (informed by the Game \ac{AI}) and also how to structure the dialogue with the students (informed by the Emotional Climate and the other modules).
\end{description}

While some of the aforementioned modules are standard in \ac{HRI} domains, some are specific to our scenario, namely in what refers to the interaction with multiple students. Below, we discuss in greater detail the methodological and technological considerations that drove the design of such components.

\subsection{Designing the interaction behaviors of the robot}%
\label{Subsec:Restricted-woz}
As an empathic tutor, the robot should be able to play the game and to interact in a social and empathic manner with the students, raising awareness for personal values when considering sustainability. The robot should do so by setting a good example when choosing its game actions, explaining the reason behind each play in light of its ``personal values'', namely, achieving a balanced development of the city. We note that, in the context of our scenario, no personal values are wrong or right, and the students are not expected to adopt the robot's personal values. Instead, it is simply a means to open up the discussion and raise awareness of others' perspectives, as active participation is a sign of learning in sustainable development education \cite{lave1991situated}. Thefeore, to design the social behavior of the robot, we adopted the restricted-perception \ac{WoZ} methodology detailed in \citeN{sequeira2016method}, which can be summarized in the following steps:
\begin{enumerate}
	\item We gathered data from mock-up sessions where two students collaboratively interacted with a school teacher playing \mcec{}. The goal was to gain insight in common pedagogical and empathic strategies used by real teachers in our learning activity.
	\item Data collected during the mock-ups was used to build the modules responsible for the perception, basic game behavior, and interaction behavior of the robot.
	\item We conducted Restricted-Perception \ac{WoZ} studies, in which experts remotely-controlled the robot during the interaction with students in the \mcec{} game. Unlike the standard \ac{WoZ} paradigm in \ac{HRI}, when applying a Restricted-Perception \ac{WoZ} method, the experts controlling the robot have access only to processed observations from the interaction, similar to those which will drive the robot's autonomous behavior. This means that the decision-making of the wizard will be limited to the same perceptions that the autonomous robot will have over the interaction. This paradigm allows the operator to focus on the relevant aspects of the robot's social interaction.
	\item Using the data collected during the Restricted-Perception \ac{WoZ} studies, we created the Hybrid Behavior Controller. The controller was built from two key elements: \emph{interaction strategy rules}, in which we encoded expert domain knowledge, e.g., explicit behavior patterns observed during the mock-up sessions and \ac{WoZ} studies; and a \emph{mapping function}, which identifies more complex behavior patterns discovered from the data using a \ac{ML} approach.
\end{enumerate}

\subsection{Implementing the interaction behaviors of the robot}
Although this section describes work that is not a direct contribution of this paper, it is crucial to understand the design and implementation decisions that were taken regarding the behavior of the autonomous robotic tutor used in the studies reported here. In particular, we discuss the core modules that support the robot's social and empathic interaction with multiple students in the context of our learning activity (see Fig.~\ref{fig:architecture}).

\subsubsection{Emotional Climate}
\label{Subsec:EC}
In order for the robot to interact with a group of students in an empathic and emotionally-intelligent way, it is fundamental that it is able to {\em detect} and {\em recognize the emotional state of the group}. Emotional climate is a central element in social group interactions between humans and has been studied in many group contexts. We consider emotional climate to be the valence state of a group-level emotion at a given time. Following the discussion in \citeN{alves2015ec}, in our scenario, the behavioral and emotional state of the students at any given time changes the emotional climate of the group at that time. This means that a positive emotional can be detected from the students expressing positive facial expressions, demonstrating joint attention in the educational task by looking at the table where the interaction takes place, etc. Conversely, a negative emotional is detected whenever students look away from the task and seem distracted or bored, etc. Emotional climate influences the behavior of the robotic tutor by changing the content and the way that certain utterances are performed. For example, if students are taking more time than the usual to decide what to do and a negative emotional climate valence is detected (e.g., boredom), the robot intervenes to maintain engagement by saying: \emph{``what should we do now?''} On the other hand, if a positive emotional climate valence is detected (e.g., engagement), the robot may say: \emph{``we are playing well''} as a positive reinforncement. These constitute examples of empathic behaviors designed for the robot.

\subsubsection{Game \ac{AI}}
\label{Subsec:GameAI}
The Game \ac{AI} module is detailed in \citeN{sequeira2015ai} and ensures that the robot tutor is not only able to play the game competently, but also discusses the impact of each action performed by the group in the construction of a sustainable city. The robot's game-play promotes collaboration within the group and comprises a \emph{game-playing} and a \emph{social} component. The \emph{game playing component} (planner) is designed to accommodate a specific educational strategy, e.g., if the goal is to achieve a ``balanced'' strategy, it favors actions leading to game states where all scores (environment, economy and well-being) are as high and even as possible. It also detects game situations with the potential to provide interesting pedagogical opportunities, e.g., when the level of natural resources is low it suggests game actions that spend fewer resources. The \emph{social component} uses information about recent plays to build a model of the students' game strategies. It allows the robot to intervene during and after the players' actions in the game, e.g., the robot is able to suggest more suitable alternative plays in certain situations and explain the desired effect of such decisions over the city's development. Such game-play model also allow the robot to adapt its own strategy so to follow the perceived group strategy, a fundamental aspect of its game behavior due to the collaborative nature of the activity. For example, if the students are playing in an environmentally-aware fashion, the robot's strategy will also be more environment-friendly.

\begin{minipage}{\linewidth}
\begin{lstlisting}[caption=Example of a robot behavior.,label=List:BehaviorExample]
Category: Strategy
Sub-category: Wellbeing
Behavior: 
{
	<gaze> 		game-ui 
	<animate> 	sadness 
	<animate> 	slow-eye-blink
	<speech> 	"Our population is not very happy."
	<glance>	subject-1
	<speech>	"This worries me."
	<glance> 	subject-2
	<speech> 	"What can we do?"
	<gaze> 		game-ui
}
\end{lstlisting}
\end{minipage}

\subsubsection{Student Modeling}
The Student Modeling component in Fig.~\ref{fig:architecture} follows the discussion in \citeN{jones2015empathic}. It provides \emph{task and domain knowledge}, i.e., information both on the learning activity and on the knowledge and skills that the student is expected to acquire. The Student Modeling uses the information from the User Action module to update a high-level description of each student. Such description includes the student's task performance and estimated domain knowledge. Both are stored in the Memory module to be used in a pedagogical manner during the interaction, e.g., the robot can show its support with respect to the users' difficulties, and by summarizing the main results achieved at the end of each learning session. In the long-term study, the students' performance is especially useful to revisit specific tasks within \mcec{} that were completed or not, as well as information about how they were completed. This allows the tutor to ``recall'' previous sessions, highlighting learning gains and discussing specific challenges that the students went through. It also allows the tutor to provide the group with hints on how to address such challenges, thereby adapting to their learning needs.

\subsubsection{Hybrid Behavior Controller}
\label{Subsec:Hybrid}
The interaction behaviors of the robot are governed by the Hybrid Behavior Controller module, whose design and implementation details can be found in \citeN{sequeira2016method}. The controller comprises a set of \emph{interaction strategy rules} and a \emph{mapping function}. The input of the controller is a set of \emph{perceptual features}, namely: facial features, encoding the students' expressive information; auditory features, identifying the active speaker and detecting a set of keywords spoken by the students that are relevant for the learning task; and game-related features, providing information about critical moments of the game, such as when a level changes or some resource of the city becomes scarce. All features are automatically extracted and encoded from raw data captured from microphones, cameras and other sensing devices strategically positioned in the environment. The output of the controller module is a \emph{social interaction behavior}, including all the animations, gaze functions and utterances spoken by the robot during the interaction with the students. The design of the behaviors, as discussed in \citeN{alves2014towards}, was inspired by observed teacher-students interactions during the aforementioned mock-up sessions. In addition to the dialog of the robot, each interaction behavior encodes the non-verbal behavior of the robot that was also inspired in the way real teachers and the students interacted, e.g, by shifting the robot's gaze between the game and the players in order to drive their focus of attention towards relevant aspects of the task. An example of a full behavior definition is specified in Listing~\ref{List:BehaviorExample}, designed to address a situation where the well-being of the city's population in \mcec{} was low\footnote{The list of encoded behaviors can be retrieved from ``Publications / Deliverables'' sections of the project's website at: \url{http://www.emote-project.eu/}.}. The Hybrid Behavior Controller is then comprised of:

\begin{figure}[!tb]
   \centering
   \includegraphics[width=\textwidth]{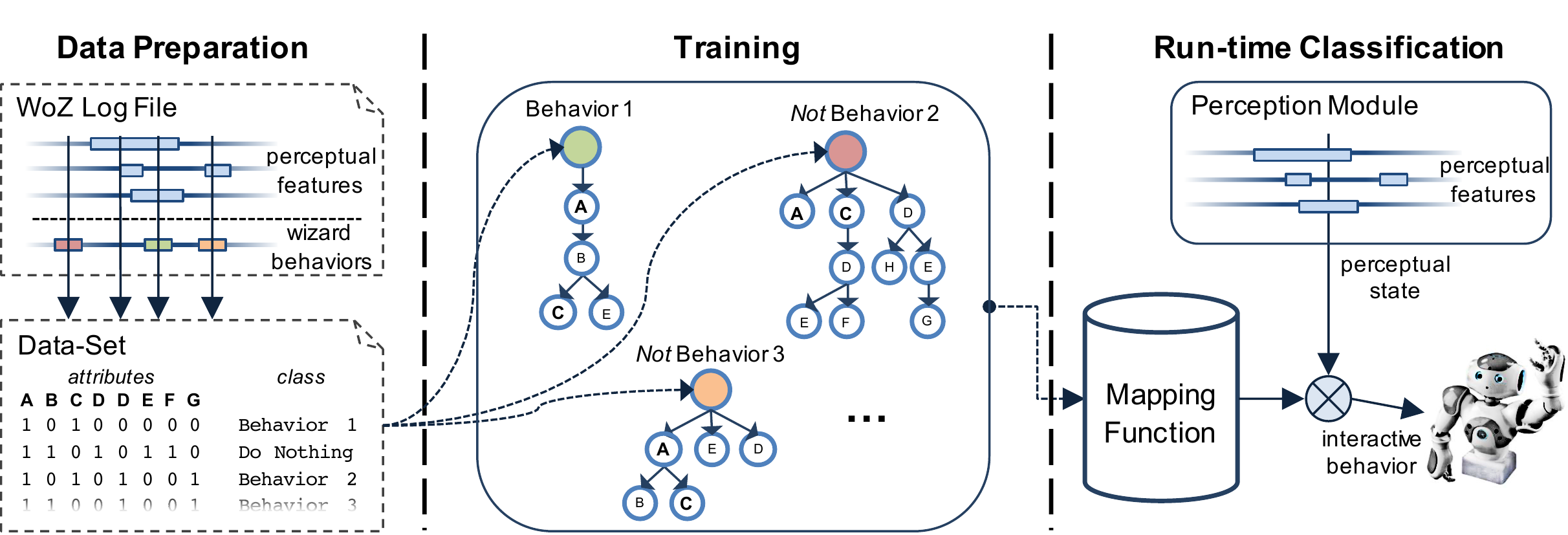}
   \caption{A depiction of the \ac{ML} procedure involved in creating the robot controller's \emph{Mapping Function}. Adapted from \citeN{sequeira2016method}.}
   \label{fig:learningMF}
\end{figure}

\begin{description}
	\item [Interaction Strategy Rules --] Correspond to manually-encoded \emph{behavior rules} in the form \texttt{If}-perceptual state-\texttt{Then}-interaction behavior. The idea is that when the features have the values as specified in the rule's \texttt{If} statement, the rule becomes active, which in turn automatically triggers the associated interaction behavior of the robot. A set of rules was defined to encode domain knowledge that is relevant to improve the students' comprehension of the task and to understand their learning progress. Some rules were inspired by pedagogical strategies observed in teachers during the mock-up sessions. We also performed informal interviews with the teachers in order to understand their reasoning process and gather information about interaction dynamics and common strategies used during the several interaction studies. This lead to the design of rules such that whenever the game starts, the robot gives a short tutorial explaining the game rules, and after the game ends a rule is triggered that ``wraps-up'' by summarizing the main achievements and analyzing the group's performance. Other rules encode interaction-management functions, such as announcing the next player or other game-related information. The rationale was that the behaviors in these rules occurred at well-defined moments and in a consistent manner, hence we do not need to learn interaction strategies for such cases.
	\item [Mapping Function --] In order to endow the robot with a more robust behavioral repertoire, the hand-designed strategy rules were complemented by interaction strategies discovered using a \ac{ML} technique. In particular, we used \ac{ML} to identify behavioral patterns that are less common and, therefore, harder to explicitly identify by the experts or through observation and annotation. An important aspect of the Restricted-Perception \ac{WoZ} method is that the behavior data from the operator is dependent on the same perceptual features that will drive the behavior of the robot during autonomous interaction \cite{sequeira2016method}. For this reason, such data is particularly amenable to a \ac{ML} analysis. We used the data from these studies to train a classifier that maps perceived situations to the robot's actions, i.e., a model of \emph{which} behaviors should be triggered and \emph{when} to trigger them. The procedure is illustrated in Fig.~\ref{fig:learningMF}. It starts with a \emph{Data Preparation} phase involving the transformation of the collected demonstrations into a data-set of state features-behavior pairs, which are referred to as training instances. The \emph{Training} phase learns a mapping function encoding the observed interaction strategies from the given data-set.%
	\footnote{We note that the interaction controller module is agnostic to the \ac{ML} algorithm that is used to learn the Mapping Function. In that regard, standard \ac{ML} classification algorithms may be suitable to learn interaction strategies based on the collected \ac{WoZ} data.}
	Specifically, we used a technique based on an associative metric within frequent-pattern mining that is detailed in \citeN{sequeira2010fpm} and in \citeN{sequeira2013assoc}. As illustrated in Fig.~\ref{fig:learningMF}, for each wizard behavior sampled from the log file, the corresponding ``Behavior'' tree is updated according to the perceptual features that were active at that time. This indicates that states where those perceptions are active are an example of \emph{when} to execute the behavior. For all other behaviors, the corresponding ``Not Behavior'' trees are updated, indicating that the features are an example of \emph{when not} to execute them. By the end of training, each ``(Not) Behavior'' tree stores patterns that indicate the perceptual states on which the corresponding behavior should (not) be executed. After having learned the mapping function, the system can choose an appropriate interaction behavior at run-time upon request, given the robot's perceptual state. We note that while the Rules module covers the question regarding when and when not to execute some behavior (the rules were handcrafted to ensure this), the \ac{ML} module had to be designed such that behaviors are not automatically-triggered incorrectly and at the wrong times, which could potentially ``break'' the interaction flow between the robot and the students. 
\end{description}

\section{Method}
\label{ch:method}
This section presents the design of the method to meet the proposed goal of this work.

\subsection{Hypothesis}
\label{Sub:Hypothesis}
A few points can be highlighted to frame our studies: teachers with empathy competencies are associated with positive students' outcomes; collaborative learning environments can be more beneficial, depending on the educational topic that is being taught; social robots have been used for children in educational applications, with positive impact on students' engagement during learning. Therefore, we have formulated the following study hypothesis:\\

\noindent%
\textbf{Hypothesis 1 -- In a collaborative group learning environment an empathic robot improves students' learning outcomes.}
We have performed a between-subjects design study in which an empathic robotic tutor interacts with groups of children in a school classroom, performing a collaborative activity about sustainable development. This was a short-term study in which groups of children performed one session with the robot, distributed randomly across one of the three study conditions: (1) two children learn with an empathic robot, (2) two children learn with a non-empathic robot (3) three children learn without the presence of a robot. We hypothesize that students will have higher learning achievements in the condition in which they perform the learning activity with the empathic robot.\\

\noindent%
\textbf{Hypothesis 2 -- In a collaborative learning environment groups of children learn over time with an empathic robot.}
This hypothesis concerns a deeper understanding about empathy in robots, as it concerns a long-term study. Thus, we have performed a within-subjects design study in which groups of students interacted with the empathic robot over a period of two months (4 sessions, 1 session every other week), and have evaluated their learning outcomes. The learning content in our research is related with sustainable development curricula, a domain of knowledge that requires group discussions and understanding of others people's opinions and perspectives in order to make sustainable decisions. We hypothesize that learning gains will increase over time.

\subsection{Ethical considerations}
We developed a robotic tutor that forms a social bond with lower-secondary students in order to promote learning in a personalized way. As \citeN{fridin2014kindergarten} and \citeN{serholt2017case} describe, this entails ethical concerns specially related to long-term interactions. These ethical concerns include attachment to the robot, deception about the robot's abilities, and robot autonomy and authority. Regarding the attachment to the robot, it was explained to all children that participated in the study exactly how long the robotic tutor will be present in their school and when it will be removed, similar to introducing a temporary teacher. We have explained to children the robot's workings and answered any questions about it to avoid deception over the robot's abilities. In relation to the robot's authority, as children are aware of the balance between expertise and authority \cite{alves2016students}, we explained that while the robot is trying to help them to accomplish learning tasks it will not be responsible for grading and does not have the authority to keep them engaged in the task. Moreover, all participants provided written informed consent from their caregivers prior to participation and assented to participate in the study when asked before the starting of each session. The guidelines of the Declaration of Helsinki and the standards of the American Psychological Association were followed.

\subsection{Measures}
\label{ch:measures}

\begin{table}
\caption{Learning goals supported by the robotic tutor and by the \mcec{} game, matched with the measurement media used to evaluate sustainable development learning outcomes.}
  \label{tab:measure2}
  \renewcommand{\arraystretch}{1.5}
  \begin{tabular}{ L{110pt} L{180pt} L{60pt} }
    \textbf{Learning goals in sustainable education} & \textbf{Measurement media} & \textbf{Section}\\\toprule
	Factual knowledge & Multiple-choice questionnaire & Section~\ref{ch:factual_knowledge}\\
	Trade-offs and multiple perspectives & Writing assessments: (1) trade-offs were measured according to the number of options considered to solve a  sustainable problem; (2) multiple perspectives were measured according to the number of arguments. & Section~\ref{ch:tradeoffs_perpectives}\\
    Personal values & Behavioral analysis about: (1) Scores comments, (2) In-depth discussions (3) Meaningful conversations & Section~\ref{ch:personal_values}\\
    \bottomrule
  \end{tabular}
\end{table}

We have designed, developed, and evaluated two different assessment media to measure learning goals in sustainable development education. The two assessement media used were \emph{writing assignments} and \emph{behavioral analysis}. Their full description is described below and sumarized in Table~\ref{tab:measure2}\footnote{All writing assignments used in the work are made available online on Deliverable 7.2 of the \emote{} project at \url{http://www.emote-project.eu/publications/deliverables}.}.

\subsubsection{Factual knowledge measure}
\label{ch:factual_knowledge}
A multiple-choice questionnaire was designed as a measure of Factual Knowledge about energy sources. The questions were designed according to the knowledge available in the \mcec{} game by a researcher of the \emote{} project who was also a teacher in school. The multiple-choice questionnaire about sustainability was piloted to determine whether the difficulty level in the pre- and post-tests assignments was appropriate, which would mean no statistical difference between pre- and post-tests scores. The pilot test was performed with $48$ children from grades 4 and 5 (the same age-group as the target students from our main study) and the difficulty level was evaluated based on the percentage of correct answers to each question. Results from the pilot test showed no significant difference between pre-test (\emph{M} = 5.0, SD = 0.16) and the post-test (\emph{M} = 4.9, SD = 0.16), \emph{p} > 0.05, therefore showing a similar level of difficulty. Both pre- and post-tests comprised 12 items each (24 items in total).

\subsubsection{Trade-offs and multiple perspectives measure}
\label{ch:tradeoffs_perpectives}
To test students' ability to understand that there are many different perspectives to debate sustainable development, we created a writing exercise that reflects a sustainable problem in which different stances can be taken. Students were instructed to provide two types of answers: (1) chose one or more solutions to solve the problem, as a measure of Trade-offs; (2) argue for the chosen solution(s) as a measure of Multiple Perspectives. We piloted the exercises with the same $48$ children. Two researchers coded the data and the reliability score (Cohen's kappa) for the number of perspectives mentioned in the argument was .86, denoting a strong agreement between coders. Results from the pilot study indicated no significant differences between the pre- (N = 23, \emph{M} = 0.70, SD = 0.93) and post-test  (N = 25, \emph{M} = 0.52, SD = 0.77) sustainable problems, enabling their use as a measure for the study.
\subsubsection{Personal values}
\label{ch:personal_values}
Learning about sustainability is not a straightforward process. According to \citeN{fior2010children}, sustainable development education is not primarily about changing attitudes, instead \emph{``environmental learning in the presence of complexity, uncertainty, risk and necessity} [they argue] \emph{must be accepting of multiple perspectives supportive of meta-learning across perspectives, and detached from the making of decisions in its (and learners') own immediate context''} \cite{fior2010children}. Thus, instead of changing attitudes and behaviors, the learning measures were designed to capture children's awareness of different perspectives and the ever-present trade-offs around sustainability. Furthermore, according to \citeN{antle2014}, sustainability curricula for elementary schools \emph{``often focus on key concepts such as balancing conservation and consumption''} \cite[p. 37]{antle2014}, while ignoring the important role that children's personal values have in learning. They argue that sustainability education for elementary school-aged children should be modeled according to the Emergent Dialogue Model, especially in the area of digital media games such as Youtupia \cite{antle2013youtopia}. The core idea of the Emergent Dialogue Model is that children should be invited to participate in \emph{personally meaningful dialogues} during the game play. In the case of our study, we have used an autonomous robot that interacts with children as a way to foster meaningful dialogues about sustainable development during the game-play of the \mcec. Thus, Personal Values were measured by analyzing the video recording collected during the study for behavioral analysis. Using the dedicated software ELAN v4.8.1 \cite{wittenburg2006elan}, each study session was coded for behavioral analysis using the video recordings of the participants while performing the collaborative group learning activity.
We have based our coding scheme on the one created by \citeN{antle2013youtopia}, for the analysis of the Emergent Dialogue Model. We were interested in the analysis of the verbal behavior of the participants during the learning activity to be able to gain insights into their meaningful participation in discussions of personal values as a way to measure learning outcomes. The coding scheme used was the following:

\begin{itemize}
  \item \textbf{Scores comment --} Discussion or comments about the game scores. This category relates to children's comments or observations of the impact of their game actions on the game scores. It is related with an increase or decrease of the scores on any game parameters. An inclusion example would be: \emph{``We are running out of money''}.\\
  \item \textbf{In-depth discussions --} Includes events in which one or more children talk about decisions on what resources and developments to use. An in-depth event involves a sense of the world or individual values, which differs from simple preferences. It must also involve reasoning using those values, typically around trade-offs between human and natural needs. As such, statements like \emph{``I think we should have houses not trees''} is a preference and was not coded. However, statements such as \emph{``No, let's build houses instead of apartments because they use less lumber, and we can make more trees into nature reserves.''} was coded as in-depth discussion because it involves values in the context of reasoning about trade-offs related with sustainability.\\
  \item \textbf{Meaningful conversations --} Includes verbal and/or physical disagreement with another's action(s), or utterances related to the sustainability domain. Meaningful conversations require an objection or stance on an issue, and therefore presenting available options or suggestions is not considered. Meaningful conversations may result in resolution, abandonment (unresolved) or unilateral decision-making. Inclusion example: \emph{``I disagree, without industry you cannot progress.''}
\end{itemize}
\begin{figure}[!tb]
   \centering
   \includegraphics[width=0.70\textwidth]{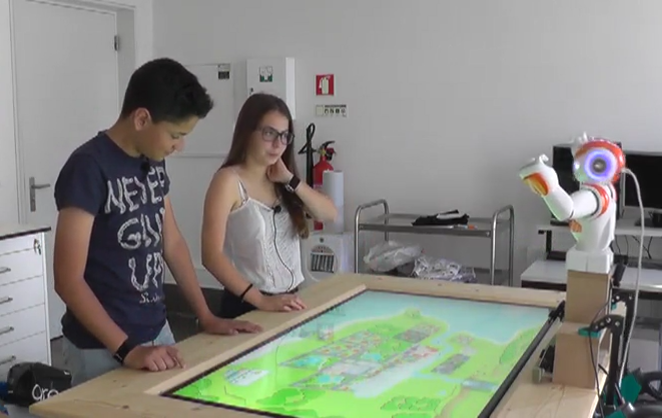}
   \caption{Two students playing the \mcec{} with our autonomous robot in a school classroom.}
   \label{fig:robot-interaction}
\end{figure}
\subsection{Materials}
The list of materials used in the set-up of both the short-term and long-term studies is listed below. Consider also Fig. \ref{fig:robot-interaction} for a picture of the real set-up.
\begin{enumerate}
\item NAO torso robot from Aldebaran Robotics;
\item Large interactive multi-touch table running the \mcec{};
\item Four video cameras for a full interaction recoding;
\item Two lavaliere microphones for voice pitch recognition (no voice recognition was used; additional details can be found in Section \ref{Subsec:Architecture});
\item Voice recorder for behavioral analysis.
\end{enumerate}

\section{Short-term study}
\label{ch:short_term_study}
An experimental study was conducted to evaluate the impact of a robot with empathy competencies on students sustainability education outcomes in a collaborative group learning environment. This relates with our Hypothesis~$1$, detailed in Section~\ref{Sub:Hypothesis}. To achieve the proposed goal, the study was designed considering three experimental conditions:\\

\begin{itemize}
\item\textbf{Condition $1$} -- Two children interacted with a robotic tutor with empathy competencies while playing the \mcec{} game.\\
\item\textbf{Condition $2$} -- Two children interacted with a robotic tutor without empathy competencies while playing the \mcec{} game.\\
\item \textbf{Condition $3$} -- Three children played the \mcec{} game without the presence of a robotic tutor.
\end{itemize}

\subsection{Empathic vs. non-empathic robot: impact on behavior}
In order to create Conditions~$1$ and $2$, featuring the robot with and without empathic competencies, we have made a choice about which modules should be activated/deactivated. We refer to Section \ref{Subsec:Architecture} for the technical architecture overview of the robotic system. Table~\ref{tab:non-empathic} details which modules from the overall system architecture, depicted in Fig.~\ref{fig:architecture}, are fully or partially (de)activated in each version of the robotic tutor. Despite some modules being deactivated or partially activated when comparing the versions of the robot, it is crucial to note that during the learning activity, the percentage of interventions by the robot toward the students was similar. This was ensured as the behaviors of the robot were designed and developed to be balanced between conditions. In practice, this means that the robot talks and gestures the same amount of time in both empathic and non-empathic versions. Additionally, basic idle behavior, animations, and speech capabilities remain intact; the non-empathic robot will, however, appear less aware of the students.
\begin{table}[!tb]
    \centering
    \caption{Overview of activation of all the modules in the empathic and non-empathic version of the robotic tutor.}%
    \label{tab:non-empathic}
    \begin{tabular}{lll}
    \multirow{2}{*}{\bf Module} 														
    & \multicolumn{2}{c}{\bf Activated} \\
    & \bf{Empathic} & \bf{Non-Empathic} \\
    \toprule
    \emph{Rapport}
    & Yes & Partially \\
    \emph{Game-\ac{AI}}
    & Yes & Yes \\
    \emph{Emotional Climate}
    & Yes & No \\
    \emph{Past Event Memory}
    & Yes & No \\
    \emph{Recent Event Memory}
    & Yes & Yes \\
    \emph{Hybrid Behavior Controller}
    & Yes & Partially \\
    \emph{Sustainable learning dialogue}
    & Yes & Yes \\
    \bottomrule
    \end{tabular}
\end{table}

As we can see from Table~\ref{tab:non-empathic}, the \emph{Past Event Memory module} is deactivated in the non-empathic condition, which means that the robot is unable to recall previous learning sessions and summarize activities that occurred therein. The rationale is that using memory of other people's past experiences is a way to simulate how they feel in situations similar to those which they are currently facing, which consequently leads to empathic behavior \cite{ciaramelli2013individualized}. Notwithstanding, the robot has the \emph{Recent Event Memory module} activated in both conditions, which means it remembers the performance of students \emph{during} each learning session in both versions.

The \emph{Emotional Climate module} is also deactivated in the non-empathic condition. This module is responsible for perceiving the emotional state of the group of students during the learning activity, an inherent perception for performing empathic behaviors. With this module deactivated in the non-empathic condition, the robot provides more generic suggestions for students during the game that are not related with emotional perceptions. Nonetheless, the \emph{Sustainable Learning Dialogue module} is activated in both empathic and non-empathic conditions and is responsible for ensuring that the robot performs similar dialogues about sustainable learning in both of the study conditions.

The \emph{Rapport module} is partially deactivated in the non-empathic condition, meaning that some contingent behaviors are still activated, but not all of them. An example that can illustrate the impact on the behavior of the robot concerns gaze, an important social signal \cite{emery2000eyes}. In the empathic condition, the robot moves his eye-gaze to the student currently speaking by accurately locating the student using the sound coming from the microphone (each student is using a microphone, as can be seen in Fig. \ref{fig:robot-interaction}, in order for the robot to accurately follow the student, however, no speech recognition system is used). In the non-empathic condition, the robot still looks at the student who is speaking, but uses predefined coordinates of the students' locations in front of the table. This means that subtle changes in the students' locations, especially of their faces, are not tracked by the robot, resulting in a less context-aware and contingency behavior towards the students. However, it is important to note that this does not translate a random gaze behavior \cite{park2017telling}, but a less precise gaze orientation that does not lead to drastic effects on the perception of the robot between conditions. Another feature of the Rapport module that is deactivated in the non-empathic condition concerns the voice volume and pitch of the robot not being adjusted to the perceived volume and pitch of the students' voices, contrarily to what occurs in the empathic condition. As these adjustments are related with empathy behaviors \cite{imel2014association}, the characteristics of the robot's voice are kept constant in the non-empathic version.

As for the \emph{Hybrid Behavior Controller module}, only the Interaction Strategy Rules are active for the non-empathic version, which means that there are no behaviors being triggered by the Mapping Function. Although this part of the controller does not necessarily lead to empathic behavior, we note that it is the result of applying a \ac{ML} algorithm aimed at discovering more subtle interaction strategies used by the wizard in the Restricted-Perception \ac{WoZ}, which cannot be hand-crafted and put in the Interaction Strategy Rules list. Notwithstanding, this does not mean that the robot will intervene inappropriately and/or at the wrong times, as all behavior is still controlled by manually encoded rules. Thus, the robot was kept as a knowledgeable and
informative interlocutor in all study conditions, as previous studies have shown that children can easily distinguish between reliable robots as information sources \cite{breazeal2016young}.

Overall, the deactivated modules concern perceptions of the cognitive and emotional states of the students. Notwithstanding, the social and pedagogical behaviors are similar in both study conditions, with the robot having similar frequency of interventions. This ensured the social and intelligent tutoring autonomous behavior of the robot.

\subsection{Sample}
A total of $63$ children ($M=13.74$, $\mathrm{SD}=0.73$, $25$ female) participated in this study. Participants were grouped by their school teachers according to groups of students they knew worked well together in a learning context, and were randomly assigned across study conditions. Therefore, $22$ participants interacted with the empathic robotic tutor, in a total of $11$ learning sessions consisting each of $2$ children and $1$ robot; $20$ participants interacted with a non-empathic robotic tutor, in a total of $10$ learning sessions consisting each of $2$ children and $1$ robot; $21$ participants were allocated in the condition with no robot, in a total of $7$ learning sessions consisting of $3$ children. Two researchers were responsible for the study in the school: a psychologist that interacted with the participants and acted as a leading researcher, and a computer scientist  that was responsible for the technical equipment.

\subsection{Procedure}
\label{ch:procedure_st}
Each group of students arrived to the designated classroom where the study took place. The leading researcher provided an explanation about the study they would undergo. Participants were invited to fill-in the pre-tests about sustainability explained in section~\ref{ch:measures}. After completing the pre-tests, participants were introduced to the robotic tutor (in case of conditions $1$ and $2$) and to the \mcec{} game, while they performed an initial trial round of the game to have a hands-on experience with the activity. When the trial round of the game was over, the researcher left the room, leaving the participants performing the learning activity with the robot (in case of conditions 1 and 2) or by themselves (in case of condition 3). Although the students were left alone in the classroom performing the learning activity, they had permanent indirect supervision by both researchers. This was ensured since the classroom had a large window to an external room, which allowed to monitor the progress and at the same time providing privacy to their learning process. Furthermore, this set-up enabled participants to reach out to the researcher to ask for help, e.g., if technical problems with the learning activity or with the robot occurred. Finally, when the learning activity was over, the researcher entered the room and closed the learning activity application, thus ending the activity. Participants were able to say goodbye to the robot and afterwards were invited to fill-in the post-tests about sustainability knowledge. At the end, some time was given to discuss their experience during the study, providing an open space for children to ask questions or share thoughts. Each session had a duration of $1$ hour, in which $30$ minutes were allocated to the activity of playing the \mcec{} game, and the remaining $30$ minutes were dedicated to pre- and post-tests.

\subsection{Results}
We present the learning gains for the different measures used about sustainable education.

\subsubsection{Factual knowledge}
We compared the results from the pre- and post-tests across the $3$ study conditions to compare the learning outcomes in the participants' factual knowledge about sustainable learning. According to the Mixed-ANOVA test, there was no significant difference between the study conditions for learning outcomes on factual knowledge about sustainable education, $F\left(2,59\right)=0.586$, $p=0.560$. The means for the pre-test result were the following: $M=9.29$, $\mathrm{SD}=1.42$; $M=9.20$, $\mathrm{SD}=1.06$; $M=9.33$, $\mathrm{SD}=1.02$, corresponding to the interaction with the empathic robot, non-empathic robot, and no-robot conditions, respectively. While the means for the post-tests were the following: $M=7.90$, $\mathrm{SD}=1.67$; $M=8.30$, $\mathrm{SD}=1.46$; $M=8.38$, $\mathrm{SD}=1.24$, corresponding to the interaction with the empathic robot, non-empathic robot, and no-robot conditions, respectively.

\subsubsection{Trade-offs and multiple perspectives}
The assessment of participants' understanding of trade-offs and perspectives was performed, as illustrated in Table ~\ref{tab:measure2}.

\paragraph*{Trade-offs (or number of solutions)} We ran a Mixed-ANOVA test to analyze if there were differences in the number of solutions chosen by the participants across the $3$ study conditions to solve the sustainable exercise problems. We took into account their performance in the pre-test compared to their performance in the post-test and the condition they were allocated in. Results showed no significant differences across conditions when comparing the results from the pre- and post-tests, $F\left(2,60\right)=1.726, \emph{p}=0.187$.

\paragraph*{Multiple perspectives (or number of arguments)} The number of arguments mentioned by the participants to justify their solutions was the measure for the multiple perspectives in solving a sustainable problem. Results show that the number of arguments did not change significantly, $F\left(2,493\right)=0.925$, $p=0.402$, when participants learned with the empathic robotic tutor compared to the other study conditions ($M=1.05$, $\mathrm{SD}=0.19$; $M=1.00$, $\mathrm{SD}=0.20$, for pre- and post-tests, respectively).

\subsubsection{Personal values}
We performed verbal behavior analysis across the $3$ study conditions to measure personal values, considering \textit{scores comment}, \textit{in-depth discussions}, and \textit{meaningful conversations} (see coding scheme in Section~\ref{ch:personal_values} for more details).
When running a Chi-squared test, we can see that there is a statistically significant association between the conditions of the study and how children exchanged personal values about sustainability, $\chi^{2}\left(4\right)=9.375$, $p=0.05$, with the strength of the relationship with Cram\'{e}r's V being $\phi_{c}=0.109$ revealing a medium effect. We performed a post-hoc test to analyze contingency tables to understand in which study conditions personal values exchanges were statistically significant \cite{beasley1995multiple}.

\begin{figure*}[!tb]
   \centering
    \begin{subfigure}{0.48\columnwidth}
   		\centering
       	\includegraphics[width=\columnwidth]{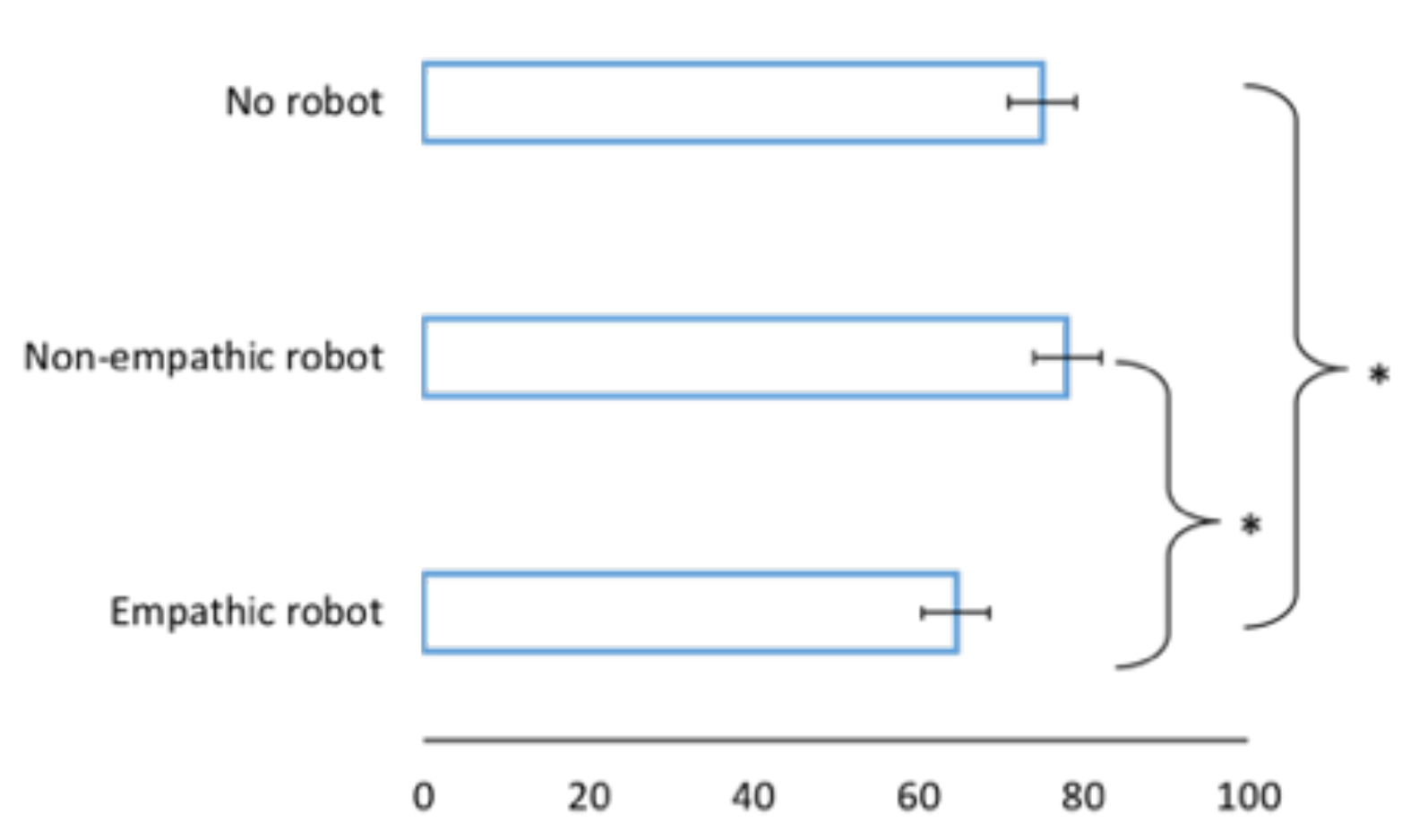}
		\caption{Score comments}\label{Fig:scores_short}
   \end{subfigure}\hspace{5pt}%
   \begin{subfigure}{0.48\columnwidth}
   		\centering
       	\includegraphics[width=\columnwidth]{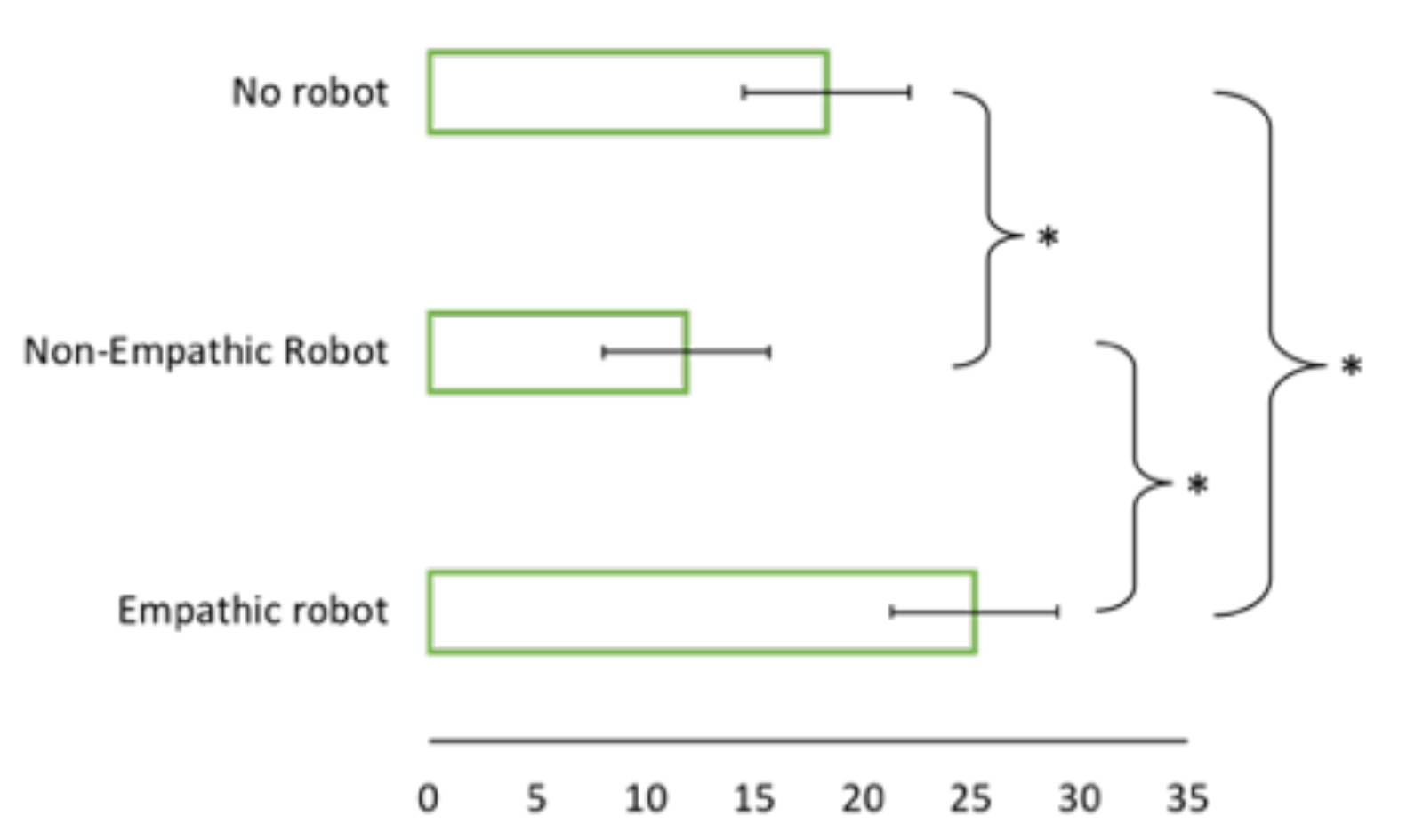}
		\caption{Meaningful conversations}\label{Fig:meaningful_short}
   \end{subfigure}
   \caption{Results for comments and meaningful conversations about sustainable development during the short-term learning activity across the 3 study conditions. Results are presented in frequencies and significant results are represented with the symbol $*$, $p<0.05$.}
   \label{Fig:st_results2}
\end{figure*}

\paragraph*{Scores comment} We found that in the empathic robotic tutor condition participants commented on scores statistically less comparing to the other study conditions, $p=0.01$. Thus, participants commented less on the scores when in the condition with the empathic robotic tutor ($64.52\%$), compared to the condition with the non-empathic robotic tutor ($77.97\%$) and with the no-robot condition ($75.00\%$), as illustrated in Fig.~\ref{Fig:scores_short}. The abovementioned results are significant for $p=0.016$, according to the procedure of residual analysis.

\paragraph*{In-depth discussions} No significant result was found for in-depth discussions between study conditions, \emph{p} > 0.05.

\paragraph*{Meaningful conversations} By using the adjusted standardized residuals method of analysis \cite{garcia2003cellwise}, we discovered significant differences in meaningful conversations between study conditions. Therefore, when children learned with the empathic robotic tutor more meaningful conversations emerged ($25.16\%$, $p=0.013$), followed by the no-robot condition ($18.33\%$, $p=0.012$), and the least meaningful discussion occurred when children learned with the non-empathic robotic tutor ($11.86\%$, $p=0.016$) (see Fig.~\ref{Fig:meaningful_short}). 

\subsection{Discussion}
This section discusses the results regarding the learning gains across the $3$ study conditions for this short-term study in a school classroom.\\

\noindent%
{\bf Children learned to have meaningful conversations about sustainability and worried less about scores when learning with an empathic robot.}

\noindent%
Because sustainable development education is about engaging in deep discussions that consider the existence of \emph{``complexity, uncertainty, risk and necessity''} to solve sustainability-related problems \cite{fior2010children}, the results showed that this was successfully accomplished when children learned with a robot with empathy competencies. However, when children interacted with a robot without empathy or without a robot at all, they seemed to be more concerned about passing levels (the traditional way of playing any game), instead of engaging in dialogue about sustainable education. We emphasize that the tutoring behaviors for the empathic robot vs. the non-empathic conditions were the same, i.e., the sustainable learning dialogue that the robots had was similar in both study conditions. This makes the result particularly important since it shows that empathic competencies in a robot impacted the way children engaged in the learning process, partially supporting our first study hypothesis. In fact, engaging in meaningful conversations implies that children share their personal values by objecting or taking a stance that generates discussion about sustainability while playing the game. The empathic robotic tutor fostered and motivated the children to engage in this type of dialogue as a way to increase their knowledge about sustainable education. Future studies should include more in depth analysis of educational interactions, specifically related with the emergence of educational dialogues (e.g., if it is facilitated by the robot or instigated by children themselves).\\

\noindent%
{\bf No other impact on sustainable development learning was found.}

\noindent%
Results from the multiple-choice questionnaire on factual knowledge about sustainability did not present significant results. Additionally, the writing exercise in which children were invited to choose solutions to solve a problem related to sustainable development and to argue about their options for solving it, also did not show significant results. The results might be a reflection of the short-term interaction with the robot. In fact, learning takes time \cite{fisher1981teaching}, especially when we consider sustainable education, a challenging curricular topic to teach that is complex to learn \cite{moore2005barriers}. The short-term interaction that students had with the \mcec{} game, which was their learning environment, can also help explain the lack of learning gains. Indeed, \mcec{} enables students to explore the virtual world of the game in an unrestricted way. By providing freedom to open game menus that seem interesting to them according to the game action they want to perform, this can result in students not opening all of the game menus; thus, they may be exposed to only part of the overall knowledge that the game can offer. Therefore, due to the short-term nature of the interaction with the learning environment and with the robotic tutor, children would benefit for more extended period of interaction to be exposed to more learning content. This aspect will be explored in the long-term study described in section \ref{ch:long_term_study}.

\section{Long-term study}
\label{ch:long_term_study}

A descriptive long-term study was conducted in order to investigate the learning outcomes of students that learned in groups with an empathic robotic tutor over an extended period of time in school. To achieve our goal, we have deployed a robot with empathy capabilities for 2 consecutive months in a school setting (4 sessions, 1 session every other week), to teach small groups of students about sustainable education using \mcec{} as the collaborative learning environment. To sustain the achievements of children within these weeks, the robotic tutor would recall leanings and difficulties of previous sessions upon starting each learning session, thus ensuring reflection over previous acquisitions. This study related wit hypothesis number $2$, explained in section \ref{Sub:Hypothesis}.

\subsection{Sample}
A total of 20 children ($M=13.78$, $\mathrm{SD}=0.65$, $9$ female) participated in the study. Due to technical problems, one session was excluded and the final sample resulted in $18$ children. The results presented for this study exclude the session with technical problems.

\subsection{Procedure}
Although this was a different study, the procedure is similar to the one present in section~\ref{ch:procedure_st}. We present in this section the variations in the procedure. Thus, participants filled-in tests about sustainable education at three time periods: (1) \emph{baseline}, to measure their initial knowledge on sustainable development before interacting with the robot and with the learning environment; (2) at the \emph{end of the first collaborative session}, to measure learning achievements after one interaction with the robotic tutor, which is typical for many studies in the \ac{HRI} field; (3) at the \emph{end of the 2-month period}, to be able to compare learning achievements and understand the learning curve after a long-term interaction with the empathic robotic tutor. Learning sessions were performed once every other week, therefore, two sessions per month in a total of four sessions in two consecutive months. Each session lasted about $30$ minutes with the first and last sessions taking longer as the assessments of sustainable education were applied in these sessions. The session dynamics were organized with the school teachers in order to minimize disturbances in the usual daily activities that children are involved in while in school.

\begin{figure*}[!tb]
   \centering
    \begin{subfigure}{0.4\columnwidth}
   		\centering
       	\includegraphics[width=\columnwidth]{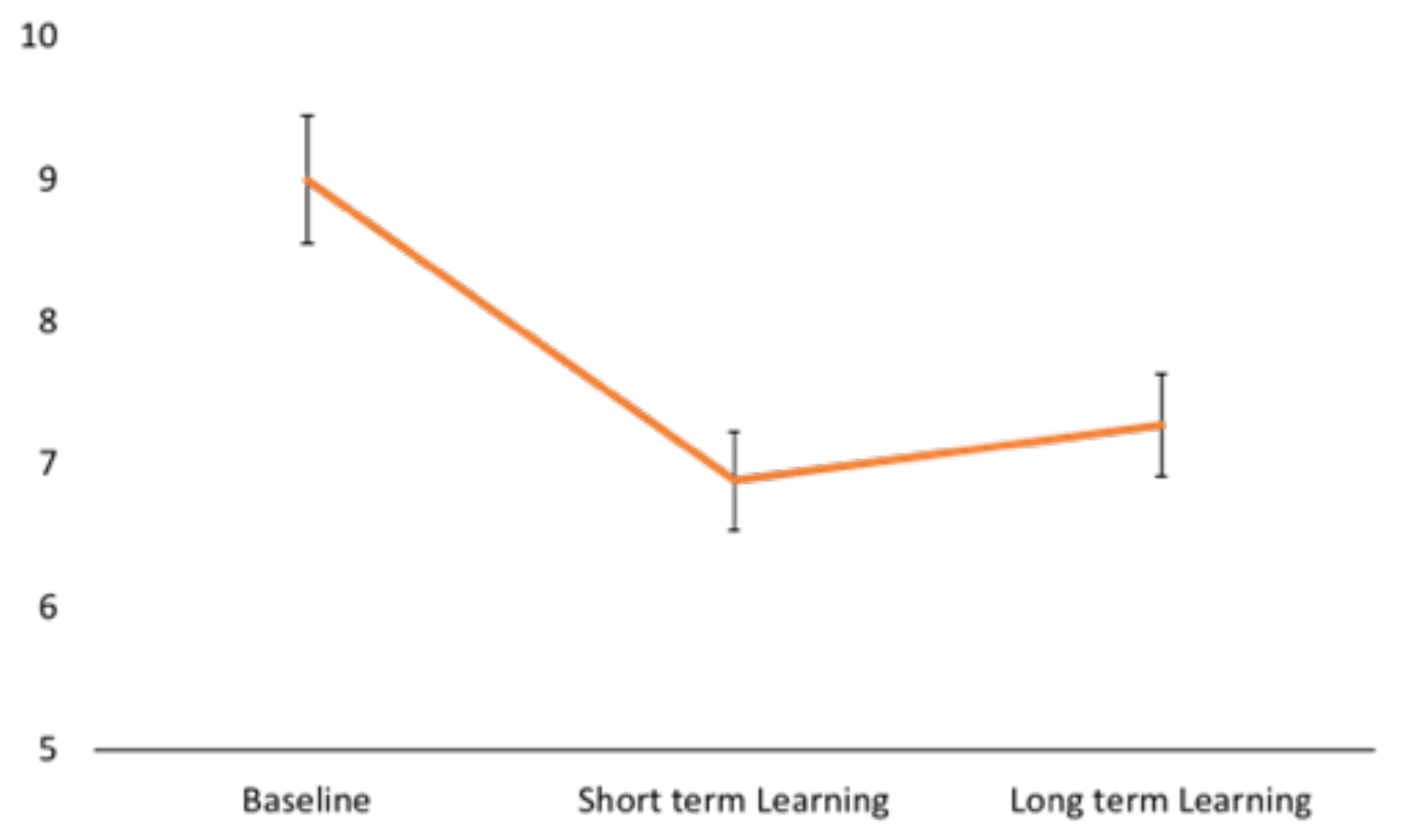}
		\caption{}\label{Fig:facts}
   \end{subfigure}\hspace{20pt}
   \begin{subfigure}{0.4\columnwidth}
   		\centering
       	\includegraphics[width=\columnwidth]{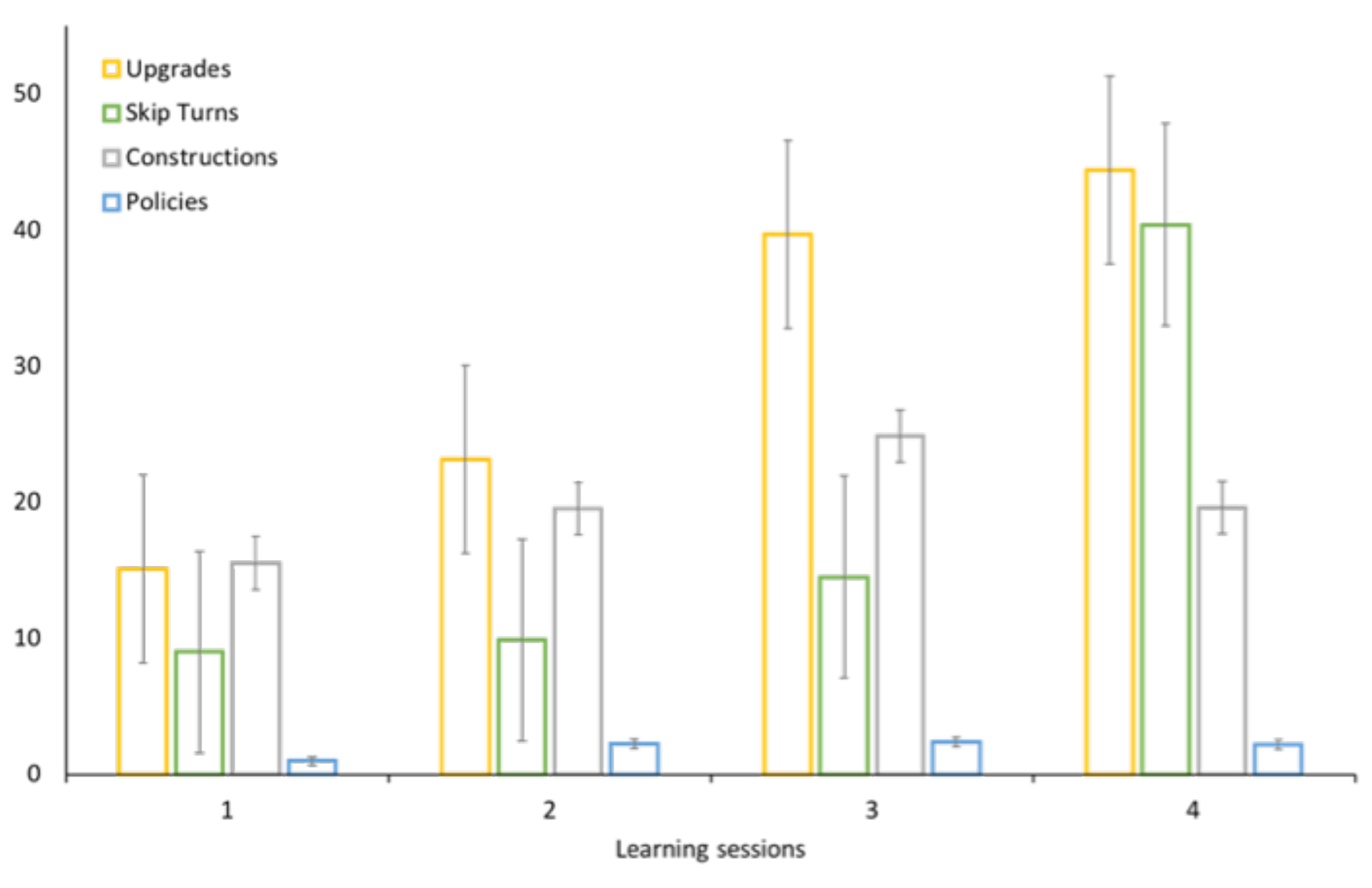}
		\caption{}\label{Fig:logs}
   \end{subfigure}
   \caption{Results of the long-term study: \subref{Fig:facts} factual knowledge achievements; \subref{Fig:facts} actions performed by students in the learning environment across the $4$ learning sessions with the empathic robotic tutor. Results revealed to be significant for the upgrades and skip turns' actions.}
   \label{Fig:long-term}
\end{figure*}

\subsection{Results}
In this Section we present the results for the long-term study with a robot with empathy in a classroom environment in school.

\subsubsection{Factual knowledge}
The factual knowledge learning about sustainability was analyzed using the Friedman's test, and students' achievements in sustainability education showed a statistically significant difference over time, $\chi^2\left(2\right)=15.464$, $p<0.001$ with a Kendall W of $.43$ indicating a moderate effect \cite{tomczak2014need}. Post hoc analysis with Wilcoxon signed-rank tests was conducted with a Bonferroni correction applied, revealing a significance difference when comparing the baseline ($M=9.00$, $\mathrm{SD}=1.41$) to the short-term learning results ($M=6.89$, $\mathrm{SD}=1.28$), $Z=-3.344$, $p=0.001$, $r=-0.56$; and when comparing the baseline with the long-term learning results ($M=7.28$, $\mathrm{SD}=1.23$), $Z=-3.084$, $p=0.002$, $r=-0.51$. No other statistical significant result was found, $p>0.05$.  From Fig.~\ref{Fig:facts}, we can see that participants' knowledge about sustainability topics started high, and after interacting with the robot in the pedagogical activity it decreased, albeit with a slightly increase in the long-term, possibly showing a tendency to return to the baseline. This result possibly translates normative results in children's learning, in which they question previously-accommodated knowledge about the topics.

\subsubsection{Trade-offs and multiple perspectives}
Similar to the analysis performed for the short-term evaluation, we evaluated the trade-offs and multiple perspectives about sustainability according to the \emph{number of solutions} proposed to solve a given sustainable dilemma, and the \emph{number of arguments} considered to justify their solutions. We explain the results below.

\paragraph*{Trade-offs (or number of solutions)}
We analyzed the number of solutions that participants considered possible to solve the given environmental problem. Participants considered more options over time, however this increase was not significant, $\chi^2\left(2\right)=1.125$, $p>0.05$.

\paragraph*{Multiple perspectives (or number of arguments)} Although there was a slight increase in the number of perspectives considered by participants, Friedman's test showed that this was not statistically significant, $\chi^2\left(2\right)=3.375$, $p>0.05$.

\subsubsection{Personal values}
A Repeated Measures ANOVA with a Greenhouse-Geisser correction determined that mean personal values did not differ significantly between the first and the last learning session with the empathic robotic tutor, $F\left(1,1.756\right)=3.530, \emph{p}=0.061$. Since personal values were measured using behavioral analysis of the $4$ learning sessions, we have no baseline result (baseline considers assessments conducted prior to the starting of the intervention).

\begin{figure}[!tb]
	\centering
	\includegraphics[width=0.7\columnwidth]{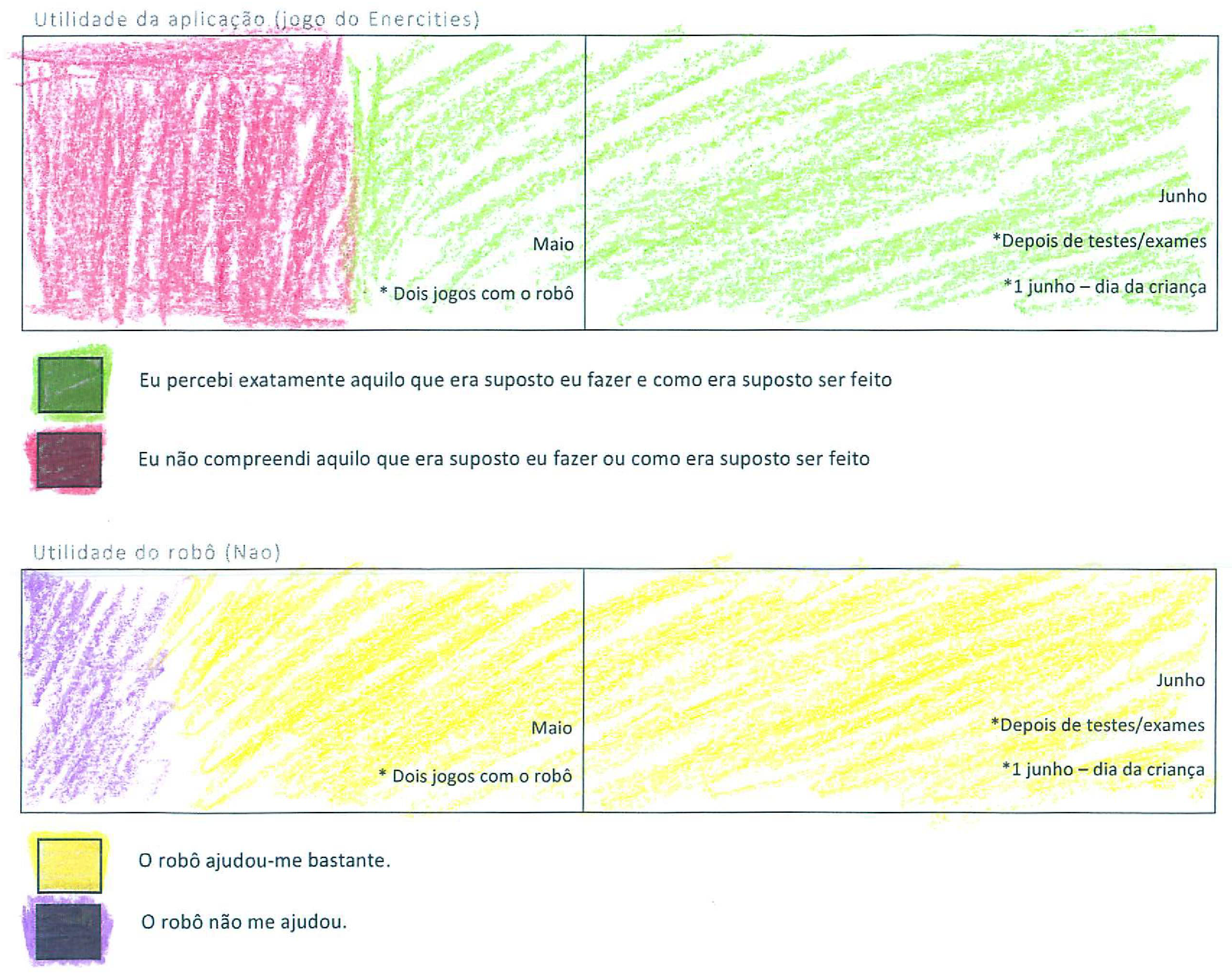}
	\caption{Example of a MemoLine filled out by one of the children and adapted for the context of the present study. Children were asked to use pencils with different colors to express how hard/easy it was for them to (1) play the game (red color signals ``I did not understand what I should do in the game'' and green signals ``I understood what I should do in the game''), (2) and how much the robotic tutor was useful for them in understanding how to play the game (purple signals ``The robotic tutor did not help me understand the game'' and yellow signals ``The robotic tutor helped me understand how to play the game''). The MemoLine should be read in timeline manner. Therefore, the line in the center signals the middle point of the sessions in order to situate children in their assessment. The left side of the line concerns Sessions~1 and 2 and the left side concerns the Sessions~3 and 4.}
   	\label{Fig:memo}
\end{figure}

\subsubsection{Game-play}
The behavior (or actions) of the participants during game-play also served as a measure for analyzing sustainability learning. By using the game logs we were able to extract what actions children performed during the game, representing where they invested and the dynamics of their game-play. The exact McNemar's test showed that participants' actions related with applying policies, $p=1.00$, and performing constructions, $p=0.125$, did not differ statistically between the first and the last session. However, participants performed statistically significantly more constructions' upgrades from the first to the last session, $p<0.001$, as they also skipped more turns (representing passage of time in the game as a way to enjoy from the city resources), $p<0.001$ (see Figure~\ref{Fig:logs}). This means that the proportion of upgrades and skipped turns was higher in the the last session compared with the first learning session.

In order to better understand the changes in game-play behavior, we analyzed how helpful the robot was towards a better understanding of how to play \mcec{}. For this, we have used MemoLine, a retrospect evaluation method dedicated to children that asks them to recall previous experiences with a given product or application \cite{vissers2013memoline}. See Fig. ~\ref{Fig:memo} for an example of a MemoLine filled in by a participant. A MemoLine was given to all participants at the end of the long-term study. According to its developers, MemoLine should be evaluated by making comparisons of the areas colored by children in a timeline perspective, in order to extract patterns of use. Our results indicated consistency in responses, as most MemoLines showed that children were slightly confused by the game and did not find the robotic tutor particularly helpful during the initial session. However, as time passed, they have perceived more help from the robotic tutor, and could better understand the game itself revealing a mastery in game-play. This result is representative of the overall sample of the study, since 18 children out of 20 filled the MemoLine in the same way. Taken together, participants' change in the behaviors of the game seem to be in line with their better understanding of the game dynamics, an understanding that was guided by the interaction with the robotic tutor. More importantly, understanding the game requires mastering concepts of sustainable education, which can explain the changes in game-play.

\subsection{Discussion}
This Section presets the discussion of the learning gains across a long-term study of $4$ sessions distributed over a period of $2$ consecutive months in school.\\

\noindent%
{\bf No learning gains were found after children interacted with the empathic robotic tutor for long periods of time.}

\noindent%
The overall results from the long-term study seems to show that although some variations in the learning outcomes of students occurred, those differences were not deemed significant after repeated educational interactions with the empathic robotic tutor. Although research about null results in learning gains with technology use are still scarce, a meta-analysis conducted by \citeN{wouters2013meta} studied factors that impacted learning gains when students used serious games for learning. The authors concluded that more learning occurred if: (1) learning sessions were complemented with other instruction methods, (2) when multiple training sessions were involved, (3) and if learned worked in groups \cite{wouters2013meta}. Althought our study involved multiple learning sessions and occurred in a group educational context, the learning sessions with the robotic tutor were not supplemented with other instruction methods. This deserves further investigation to understand the role of robots in learning contexts and their long-term effects on students' learning gains.

Similarly, \citeN{hew2007integrating} discussed factors related with successful technology integration in schools, such as the reconsideration of assessment metrics. The authors refer that \textit{``because curriculum and assessment are closely intertwined, there is a need to either completely reconsider the assessment approaches when technology is integrated into the school curriculum, or consider more carefully how the use of technology can meet the demands of standards-based accountability''}. This suggests that metrics used to measure students' learning gains that account for technology inclusion (e.g., robots) require further investigation to meet curricular goals. In the case of our study, despite the support provided to students, both from the robot and stimulated by the \mcec{}, it did not impacted in a measurable way students' learning in sustainability education present in formal school curriculum.

Another explanation for the lack of learning progress is provided by \citeN{gerber2001teacher} and concerns the lack of clearly defined roles of educational aids that can hinder learning gains due to an undefined presence in school. A result presented in a previous publication related with our learning scenario, showed that students assigned the role of a classmate to the robot although being explicitly instructed they would be learning with a robotic tutor \cite{alves2016role}. Therefore, we can be a facing a need to develop precise design guidelines to develop specific roles for robots in the educational sector whose goal is to increase learning outcomes.

In addition to this, more research is needed when designing and including robots for education. For example, studies have been showing that the mere physical presence of a robot can lead to learning gains \cite{leyzberg2012physical, kennedy2015robot}, however the variables that affect learning gains are not established yet. Furthermore, the design of robots for education should be tailored to the time required to learn certain curricular concepts, and we hypothesize that sustainable learning can be a case in which more learning time is required. The null results from our long-term study seem to indicate that long-term \ac{HRI} installations for education require more investigation and may even require a change in the interaction design between the robot and children. Our work introduced this discussion highlighting the need for a better understanding of long-term deployment of social robots amonst the educational sector.\\

\noindent%
{\bf Game-play behavior changes over time due to a better understating of the game guided by the robotic tutor.}

\noindent%
The way children played the game about sustainable education changed over time and this change seems to be related with the interaction with the robotic tutor. Indeed, there was a statistically significant result found in the game-play behavior of children during the long-term study. Children's actions in the game showed that, over time, they have performed more upgrades to their city they have performed more skip turns (representing the passage of time in the game, such as days passing by). By performing upgrades, the city can become more sustainable, and by skipping some turns the players allow the structures and upgrades they have chosen to implement, to get their full effect before advancing to the next level. This behaviour thus seems to indicate a more thoughtful design of the city, which matches with Antle's design principles for sustainability games \cite{antle2014}. These changes in game play are not trivial, as children needed to move away from the traditional competitiveness of passing levels (a so-called traditional mind set of a game-play), to become concerned about spending less money and taking more advantages of the resources from previous constructions. This seems to show that the change in game-play behavior goes hand in hand with the perceived easiness to understand and play the the game, guided by the interactions with the robotic tutor.

\section{General discussion and conclusions}
\label{ch:conclusion}

In this paper, we presented a novel educational scenario for social robots, in which a group of children interacts with an autonomous robot in a serious collaborative game. The goal of the interaction was to infer learning outcomes with regards to environmental sustainability and the associated trade-offs involved when designing a city. We conducted a short-term study that addressed the effects of the empathic robot and a long-term study that addressed the long-term deployment of the robot in school. The short-term study compared three conditions: empathic robot, non-empathic robot and no robot. The results showed no significant results in the majority of learning outcomes; however, there was an increase in meaningful conversations with the empathic robotic tutor, which is a stated goal for collaborative learning scenarios targeting sustainability. During the long-term study, the empathic robot was deployed for $2$ months in school. Results showed no significant change in learning gains over time. Additionally, a change in game-play behaviors related to the game was observed, in which children perform more game actions towards sustainability over the sessions. The lack in learning progress may be due to several aspects, such as the quality of the interaction, the role of robots in school, and group dynamics. This reflects the need to conduct additional research in group interactions between humans and robots for educational purposes. 

Summarizing, the highlights of our research are:

\begin{enumerate}
    \item We designed, developed, and evaluated a robot tutor for education that autonomously interacted with students in a real-world environment of a school for $2$ months.
    \item We contribute to the field of group interaction studies between humans and robots by framing the educational context as a collaborative group learning scenario.
    \item We concluded that an autonomous robot with empathy (compared with a robot without empathy or learning without a robot) fosters meaningful discussions about sustainability between students, which is a learning outcome in sustainability education.
    \item We concluded that long-term educational interactions with an empathic robot did not impacted in a significant way learning gains, although there was a change in game-actions to achieve more sustainability during game-play.
\end{enumerate}

\subsection{Limitations and future work}
Empathy is a complex construct that is highly dependent on the content of what people say to each other. Our empathic robot was able to perform contingent behaviors that translate empathy by receiving limited input from the children, but had no access to their verbal discussions. For effective empathic robotic tutors to operate in group learning environments, they should be able to understand what children say and accordingly personalize their empathic responses. To this end, developments in speech recognition for children are needed to build better empathic interactions. 

We also did not observe as many learning gains as expected, highlighting the importance of conducting more research in collaborative group learning environments with robots. Aspects such as the time duration for a learning gain to occur need to be considered when deploying a robot in a school setting. Additional qualitative research is also needed to understand the factors that favor learning gains and the factors that can hinder it. 

Regarding the wider use of robotic tutors in learning, there is already a large body of research investigating robots for second language acquisition, handwriting skills, and even the understanding of other complex curricular topics, such as chemistry and wind formations. Although this reflects positive and promising directions for \ac{HRI} for education, more educational scenarios need to be considered to understand how robots can be used for best impact on learning. This work has provided a corpus of reflection for future research, leading to questions, such as ``which variables lead to learning gains when using a robot for collaborative group education?'' and ``what is the role that a robot can have in school that fosters learning gains?'' With this work, we have begun to explore the potential for robots in group learning, bringing attention to empathy as an important competence for a robot to have when interacting with students.

\section*{Acknowledgements}
This work was partially supported by national funds through Funda\c{c}\~{a}o para a Ci\^{e}ncia e a Tecnologia (FCT) with reference UID/CEC/50021/2013, through the project AMIGOS grant n. PTDC/EEISII/7174/2014, the Carnegie Mellon Portugal Program and its Information and Communications Technologies Institute, under project CMUP-ERI/HCI/0051/2013, and by the EU-FP7 project EMOTE under grant agreement no. 317923. P. Alves-Oliveira acknowledges a FCT PhD grant ref. SFRH/BD/110223/2015. We show our gratitude to the teachers, students, and school-staff from Escola Quinta do Marqu\^{e}s (Oeiras, Portugal) for their involvement in the studies. We also thank Daniel Silva for collaborating in the coding of the verbal behavioral. The authors are solely responsible for the content of this publication. It does not represent the opinion of the European Commission (EC), and the EC is not responsible for any use that might be made of data appearing therein.

\bibliographystyle{ACM-Reference-Format}
\bibliography{tjhri_bibliography}

\end{document}